\documentclass[aps,preprint,epsfig,rotate]{revtex4}
\usepackage{graphicx}
\usepackage{epsfig}
\usepackage{amsmath}
\begin{document}
\newcommand{\pr}{\partial}
\newcommand{\rta}{\rightarrow}
\newcommand{\lta}{\leftarrow}
\newcommand{\ep}{\epsilon}
\newcommand{\ve}{\varepsilon}
\newcommand{\p}{\prime}
\newcommand{\om}{\omega}
\newcommand{\ra}{\rangle}
\newcommand{\la}{\langle}
\newcommand{\td}{\tilde}

\newcommand{\mo}{\mathcal{O}}
\newcommand{\ml}{\mathcal{L}}
\newcommand{\mathp}{\mathcal{P}}
\newcommand{\mq}{\mathcal{Q}}
\newcommand{\ms}{\mathcal{S}}

\newcommand{\nl}{$\newline$}
\newcommand{\nll}{$\newline\newline$}

\title{Non-Markovian Second-Order Quantum Master Equation
 and Its Markovian Limit:  Electronic Energy Transfer in Model Photosynthetic Systems}


\author{Navinder Singh$^{1,2}$ and Paul Brumer$^1$}
\affiliation {$^1$Chemical Physics Theory Group, Department of
Chemistry, University of Toronto,  Toronto, Ontario M5S 3H6,
Canada}\affiliation {$^2$Physical Research Laboratory,
Navrangpura, Ahmedabad-380009, India. }
\begin{abstract}
A direct numerical algorithm for solving the time-nonlocal non-Markovian
master equation in the second Born approximation 
is introduced and the range of utility of this approximation, 
and of the Markov approximation, is analyzed for the traditional 
dimer system that models excitation energy transfer
in photosynthesis.
Specifically, the coupled integro-differential
equations for the reduced density matrix are
solved by an efficient auxiliary function method in both the
energy and site representations.
In addition to giving exact results to this order, the approach
allows us to computationally assess the range of the reorganization energy and
decay rates of the phonon auto-correlation function for which
the Markovian Redfield theory and the second order
approximation is valid.
For example, the use of Redfield theory
for $\lambda> 10~\textrm{cm}^{-1}$ in systems like Fenna-Mathews-Olson (FMO) type
systems is shown to be in error. 
In addition,  
analytic inequalities are obtained for the regime of validity of the Markov
approximation in cases of weak and strong resonance coupling, allowing for 
a quick determination of  the utility of the Markovian dynamics in parameter
regions. Finally, results for
the evolution of states in a dimer system, with and without initial
coherence, are compared in order to assess the role of initial coherences.
\end{abstract}
\maketitle


\section{Introduction}

The quantum mechanics of open systems, i.e. systems interacting with an
external environment, is currently the focus of widespread 
attention\cite{kuhn,hpb,schlosshauer}.
Of particular recent interest is the issue of the extent to which quantum
coherence of the system is maintained in the presence of the environment.
Two significant modern examples may be noted: (a) the need to maintain coherence
in order to implement methods for  quantum mechanically
controlling molecular processes\cite{ourbook},  and (b) issues of the role of such
quantum coherent processes in natural environments, such as the observed long-lived 
coherent electronic energy transfer (EET) in photosynthesis\cite{gs}.

In either of these cases, and in many other examples as well, dynamical
evolution of the open system provide a significant computational challenge.
As such, a variety of approximations are often invoked to propagate the
system, such as the second Born approximation to master equations and the
Markov approximation, both of which are the focus of this paper.
In particular, 
in this paper we introduce a simple method to solve the
second Born quantum master equation
without doing the Markov approximation on
the slowly decaying envelope of the density matrix. In addition to being 
straightforward, 
this approach allows, by comparing 
to results using the Markov approximation,
a reliable determination of the range of
coupling strengths  and decay rates of the bath
auto-correlation function, over which one can use the Markovian
theory and the second order approximation.  In addition,
by examining the size of the fourth order term, this approach affords
an estimate of the range of validity of the widely used 
second-order approximation for model photosynthetic systems.

Although  the  method  developed here is applicable to general
systems with exponential bath correlation functions,
for computational simplicity we study, as do others, the dimer
system, generally regarded as a simple photosynthesis EET model.
In this case our Markovian analysis contrasts
with, for example, that in Ref. \cite{if} in which Markovian
Redfield theory is used apriori and its consequences analyzed, as opposed
to comparison with exact results.

The particular challenge arising from these EET types of systems relates to
the parameter range in which the dynamics occurs. Specifically,
quantum dynamics can be readily analyzed in two
limiting cases defined by  the relative contributions of the
inter- system
coupling  $V$ responsible for excitation energy transfer (EET),
and system-bath coupling constant $\alpha_{SB}$, responsible for
decoherence. These parameters define  two important time scales:
the excitation transfer time scale $\tau_{transfer}\equiv
\hbar/V$, and the decoherence time scale $\tau_{deco} \equiv
\hbar/\alpha_{SB}$.
If the system-bath coupling is very weak and $\tau_{deco} \gg
\tau_{transfer}$, the system is almost closed and  the
Schr\"{o}dinger equation can be used to study the dynamics.  In
the opposite case $\tau_{deco} \ll \tau_{transfer}$ (strong
system-bath coupling), the system is  open, the decoherence rate
is very fast, the dynamics is almost incoherent and a
simple Pauli type master equation description suffices. These
limiting regimes are well understood. Many real systems, such as
a number of harvesting systems \cite{tf,books, kn, kk, r1, r2, r3, r4, r5, r6, r7, r8}, however, fall between these extremes.  Recent observations \cite{gs}   of the long-lived EET has reactivated interest in these systems.

The standard approach  used to treat this intermediate regime
is to use the second Born quantum master equation \cite{kuhn},
a perturbative  master equation up to second order in
system-bath interaction  with weak system-bath coupling, plus its
Markovian approximation (e.g., as in the Redfield master
equation). Recently, two approaches have been studied for
arbitrary coupling regimes. One is based on weakening the
system-bath coupling removal  of system-bath
interaction and repartitioning the Hamiltonian term using a
polaron transformation, followed by the standard second Born master
equation \cite{jang}. The second approach is based on a reduced
hierarchy equation of Kubo and Tanimura, starting
from the path integral approach for quantum dissipative
systems \cite{if2}.  Additional methods are also being 
developed by the community.  
Here, as noted above, we introduce and utilize a particularly direct approach.

In the following section (Section II) we outline the basic model for a
dimer. In Section III we introduce the second Born quantum master
equation and phonon correlation function, diagonalize the
Hamiltonian  and  cast the master equation into  both the site and
energy representations. Section IV gives a new
auxiliary function method of solving these equations,
and provides an analysis of results  in the  Markovian
approximation.
Discussion of the results and the underlying physical
picture is given  at the end of Section V. In Section VI we
discuss the regime of validity of the second order approximation
in the master equation by estimating the order of magnitude of the
fourth order term.

The vast majority of treatments in the literature utilize coherence-free
initial conditions and study the subsequent dynamics. Results for
these initial conditions are compared to that obtained with model excitation with
weak light in Section VII. The last section provides a  brief conclusion.

\section{The Model: Dimer System}
Consider a model dimer system given by the following standard Frenkel
exciton Hamiltonian \cite{if}:

\begin{eqnarray}
H_{tot}& =& H^{el} +H^{reorg} + H^{ph} +H^{el-ph} \\
H^{el}&= &\sum_{n=1}^2 \ep_n^0 |n\ra \la n| + J (|1\ra\la2|
+|2\ra\la 1|)\\
H^{reorg}&=& \sum_{n=1}^2 \lambda_n |n\ra \la n|,~~~ \lambda_n =
\sum_i \hbar \om_i d_{ni}^2/2\\
H^{ph}&=& \sum_{n=1}^2  h_n^{ph} ,~~~~ h_n^{ph} = \sum_i \hbar
\om_i (p_i^2 +q_i^2)/2 \\
H^{el-ph} &=&  \sum_{n=1}^2 V_n u_n,~~~~~~ V_n = | n\ra \la n|,
~~~ u_n = - \sum_i \hbar \om_i d_{ni}q_i
\end{eqnarray}
Here $| n\ra$ represents the state in which only the $n^{th}$ site
is excited and all others are in the ground state. The quantity
$\ep_n^0$ is the excited electronic energy of the $n^{th}$ site in
the absence of phonons, and $J$ is the electronic coupling between
the sites which is responsible for EET. The ground state energies
of the donor and acceptor are set equal to zero and $\lambda_j$ is
the reorganization energy of the $j^{th}$ site that is dissipated
in the bath after the electronic transition occurs. The quantity
$d_{ji}$  is the dimensionless displacement of the equilibrium
configuration of the $i^{th}$ phonon mode between the ground and
the excited electronic state of the $j^{th}$ site, and $q_{i}, p_i$
are the dimensionless coordinates and momenta of the $i^{th}$ phonon mode
of frequency $\omega_i$.

\section{The second-Born quantum master equation}

The method of projection operators used to obtain open system master equations
is well known \cite{hpb}.
With the help of projection operators one can obtain the
following quantum master equation for the reduced density
matrix of the system in the second Born approximation,
which is valid when system-bath coupling is weak as compared
to the characteristic energy scale of the system [see, e.g., Ref.
\cite{kuhn}].
\begin{eqnarray}
\frac{\pr \rho^{I}(t)}{\pr t}= &-&\frac{i}{\hbar} \sum_{j=1}^2\la
u_j \ra [V_j^{I},\rho^{I}] - \frac{1}{\hbar^2} \sum_{i,j=1}^2
\int_0^t d\tau \nonumber \\ &&\left(C_{ij}(t-\tau)
[V_i^I(t),V_j^I(\tau) \rho^I(\tau)] - C_{ij}^*(t-\tau)
[V_i^I(t),\rho^I(\tau) V_j^I(\tau)] \right)\label{eq6}
\end{eqnarray}
Here, the interaction representation has been  used, which is
defined for system operators as,
\begin{equation}
\hat{O}^I(t) = U^\dagger_S (t) \hat{O} U_S(t)
\end{equation}
where $U_S(t) = \exp(-\frac{i}{\hbar} \hat{H}_s t)$ is the time
evolution operator, and $\hat{H}_s = \sum_{n=1}^2 (\ep_n^0
+\lambda_n) |n\ra \la n| + J (|1\ra\la2| +|2\ra\la 1|)$ is the
system Hamiltonian. Here, the bath is assumed to be a continuum of
harmonic oscillators, and the bath correlation functions are
defined as
\begin{equation}
C_{ij}(t)\equiv \la u_i(t)u_j(0)\ra - \la u_i\ra \la u_j\ra~.
\end{equation}
Below, the canonical average of the bath operators, $\la u_j \ra$,
 which involve the averaging over the product of displacement
and bath position co-ordinates is taken to be zero.
 The above master equation [Eq. (6)] is also termed the {\it time
convolution equation} and can be obtained from the Nakajima-Zwanzig
equation with a zeroth order approximation to the time evolution
operator in the kernel \cite{hpb}.

Converting this master equation [Eq. (6)] back to the Schr\"{o}dinger
representation gives

\begin{eqnarray}
&& \frac{\pr \rho(t)}{\pr t} = -\frac{i}{\hbar} [H_s,\rho(t)] -
\frac{1}{\hbar^2} \sum_{i,j=1}^2 \int_0^t d\tau  \label{eqnine} \\
&& \left(C_{ij}(t-\tau) [V_i(t), U_s(t-\tau) V_j\rho(\tau)
U_s^\dagger(t-\tau)] - C_{ij}^*(t-\tau) [V_i,U_s(t-\tau)\rho(\tau)
V_j U_s^\dagger(t-\tau)] \right) \nonumber.
\end{eqnarray}

We consider the case where the characteristics of the bath as seen by both
the sites are the same, and there is no bath correlation between
the sites. The  bath correlation function  is then of the form
$C_{ij}(t) = C(t) \delta_{ij}$, where
\begin{equation}
C(t) = \int_{-\infty}^{+\infty} \frac{d\omega}{2\pi} C(\omega)
e^{-i \omega t}. \label{eq10}
\end{equation}

\begin{equation}
C(\omega) = 2 \hbar (1+ n(\omega)) J(\omega), \label{eq11}
\end{equation}
where $n(\omega)$ is the Bose-Einstein distribution function.
Assuming the Drude-Lorentz model for the spectral density
$J(\omega) = 2 \lambda \frac{\omega \gamma}{\omega^2 + \gamma^2}$
where $\lambda$ is the reorganization energy,
and assuming the high temperature
approximation ($\frac{\hbar \omega}{k_B T} << 1$), as is appropriate
for the systems like the FMO model [see Refs. \cite{if} and \cite{if2}], 
we obtain the correlation function as,
\begin{equation}
C(t)= \frac{2\lambda}{\beta} e^{-\gamma t},~~~ \beta = \frac{1}{k_B T}
\label{eq12}
\end{equation}

\subsection{Explicit Site Representation of the Non-Markovian Master Equation}

For  the  explicit site representation we need the eigensystem
of the  Hamiltonian.  Assuming  $\lambda_1  =  \lambda_2 \equiv \lambda$ the
eigenvalues $E_i$ and eigenvectors $|e_i\ra$ for the system Hamiltonian
\begin{equation}
H_s = \sum_{n=1}^2 (\ep_n^0 +\lambda_n) |n\ra \la n| + J
(|1\ra\la2| +|2\ra\la 1|) \label{eq13}
\end{equation}
can be easily obtained as
\begin{eqnarray}
&&E_{1,2} = \frac{1}{2}(\ep_1^0 +\ep_2^0 + 2\lambda \mp \sqrt{(\ep_1^0 +\ep_2^0 + 2\lambda)^2 - 4 (\ep_1^0\ep_2^0 - J^2 + \lambda(\ep_1^0+\ep_2^0) + \lambda^2)})\nonumber\\
&& |e_1\ra = \frac{1}{\sqrt{\alpha_1^2 +1}}\binom{\alpha_1}{1},~~~ |e_2\ra = \frac{1}{\sqrt{\alpha_2^2 +1}}\binom{\alpha_2}{1} \nonumber\\
&&\alpha_{1,2} = \frac{1}{2 J} (\Delta \mp \sqrt{\Delta^2 + 4
J^2}),~~~\Delta = \ep_1^0-\ep_2^0. \label{eq14} \end{eqnarray}

Here the column vectors denote components in the site basis, the eigenkets are normalized and, since $\alpha_1\alpha_2 = -1$, they are orthogonal. With lengthy but
straightforward calculations, Eq. (\ref{eqnine}) for the reduced
density operator can be written explicitly in the site
representation, using Eqs. (\ref{eq10}) and (\ref{eq11}), and  as a
set of coupled integro-differential delay equations,
\begin{eqnarray}
 \frac{dx(t)}{dt} = && -2\frac{J}{\hbar} y_2(t)\nonumber\\
\frac{dy_1(t)}{dt}= &&\frac{\Delta}{\hbar}y_2(t) - \frac{4\lambda}{\beta\hbar^2}e^{-\gamma t} \int_0^t d\tau
e^{\gamma \tau} \nonumber\\
&&\left[\eta_1 \cos(E_{12}(t-\tau))y_1(\tau) +\eta_2 \sin(E_{12}(t-\tau))y_2(\tau) \right]\nonumber\\
\frac{dy_2(t)}{dt}=&&-\frac{\Delta}{\hbar}y_1(t) -\frac{J}{\hbar}(1-2x(t)) - \frac{4\lambda}{\beta\hbar^2}e^{-\gamma t} \int_0^t d\tau e^{\gamma \tau}\nonumber\\
&& \left[-\eta_2 \sin(E_{12}(t-\tau))y_1(\tau) +\eta_3
\cos(E_{12}(t-\tau))y_2(\tau)+2 \Omega y_2(\tau) \right]
\label{eq15}
\end{eqnarray}
with $ \eta_1= 1,~~~\eta_2 = -\frac{\Delta}{\sqrt{\Delta^2 +4
J^2}},~~~\eta_3=\frac{\Delta^2}{\Delta^2 + 4 J^2},~~~~
E_{12}=(E_1-E_2)/\hbar = -\frac{\sqrt{\Delta^2 +4 J^2}}{\hbar},~~~\Omega=
\frac{2 J^2}{\Delta^2 + 4 J^2}$. Here $x(t)\equiv
\rho_{11}(t)\equiv \la 1 |\hat{\rho}(t)|1\ra$ (site),
$y_1(t) \equiv \textrm{Re}[\rho_{12}(t)]$, and
$y_2(t)\equiv \textrm{Im}[\rho_{12}(t)]$, with subscripts denoting the sites.

\subsection{Energy Representation of the Non-Markovian Master Equation}

The kets $|e_{1,2}\ra$ in Eq. (\ref{eq14}) are the eigenstates of
the Hamiltonian $H_s$. The equation for a general element of the
reduced density matrix in energy representation
\begin{equation}
\rho_{ab}^e(t) \equiv \la e_a|\hat{\rho}(t)|e_b\ra,
\end{equation}
is obtained from Eqs. (\ref{eqnine}) and (\ref{eq10}) as

\begin{eqnarray}
 \frac{d\rho_{ab}^e(t)}{dt} = && - i \omega_{ab}\rho_{ab}^e
- \frac{1}{\hbar^2} \sum_{i,c,d =1}^2\int_0^t d\tau C(t-\tau) \nonumber \\
&& \left[V_i^{ac}V_i^{cd}e^{-i \omega_{cb}(t-\tau)}\rho_{db}^e(\tau)- V_i^{ac}V_i^{db}e^{-i \omega_{ad}(t-\tau)}\rho_{cd}^e(\tau) \right]\nonumber\\
&&- C^*(t-\tau)\left[V_i^{ac}V_i^{db}e^{-i
\omega_{cb}(t-\tau)}\rho_{cd}^e(\tau)- V_i^{cd}V_i^{db}e^{-i
\omega_{ad}(t-\tau)}\rho_{ac}^e(\tau) \right], \label{eq17}
\end{eqnarray} with $\omega_{ab} = (E_a - E_b)/\hbar$ and
\begin{equation}
V_1^{ac} = \frac{\alpha_a\alpha_c}{\sqrt{\alpha_a^2
+1}\sqrt{\alpha_c^2+1}},~~~ V_2^{ac} = \frac{1}{\sqrt{\alpha_a^2
+1}\sqrt{\alpha_c^2+1}}. \label{eq18}
\end{equation}
Results in the energy representation, using Eq. (\ref{eq12}) are discussed below.

\section{Method of solution: Non-Markovian}

To obtain a solution for the non-Markovian case, we first
convert the coupled integro-differential equations in the site
representation [Eq. (\ref{eq15})] to a larger number of  coupled
ordinary differential equations, a transformation made possible by the
exponential form of the correlation function. The resultant
coupled ordinary differential equations can be numerically
solved easily. This
transformation is performed as follows. First, for computational
simplicity we put $\tau' = \gamma \tau$ in Eq. (\ref{eq15})  and then
$\gamma t = t'$ in the resulting equations, and define three
auxiliary functions $f_i(t')$:
\begin{eqnarray}
&& f_1(t')\equiv \int_0^{t'} d\tau' e^{\tau'} \left[ \cos[\frac{E_{12}}{\gamma}(t'-\tau')] \td{y}_1(\tau') + \eta_2 \sin[\frac{E_{12}}{\gamma}(t'-\tau')]\td{y}_2(\tau') \right],\nonumber\\
&&f_2(t')\equiv \int_0^{t'} e^{\tau'}\td{y}_2(\tau')d\tau',\nonumber\\
&&f_3(t')\equiv \int_0^{t'} d\tau' e^{\tau'} \left[ -\eta_2 \sin[\frac{E_{12}}{\gamma}(t'-\tau')] \td{y}_1(\tau') + \eta_3 \cos[\frac{E_{12}}{\gamma}(t'-\tau')]\td{y}_2(\tau') \right].
\end{eqnarray}
Here,~$\td{y}_1(t') \equiv y_1(t'/\gamma),~~ \td{y}_2(t') \equiv
y_2(t'/\gamma)$ and we also define $\td{x}(t') \equiv
x(t'/\gamma)$. We then obtain six coupled ordinary differential
equations, three from Eq. (\ref{eq15}) and three from  differentiating the
three auxiliary functions, giving:

\begin{eqnarray}
&&\dot{\td{x}}(t') = -\frac{2 J}{\gamma \hbar} \td{y}_2(t'),\nonumber\\
&& \dot{\td{y}}_1(t') = \frac{\Delta}{\gamma \hbar}\td{y}_2(t') - \frac{4 \lambda}{\beta \gamma^2 \hbar^2} e^{-t'} f_1(t'),\nonumber\\
&& \dot{\td{y}}_2(t') = -\frac{\Delta}{\gamma \hbar}\td{y}_1(t')-\frac{J}{\gamma \hbar} + 2\frac{J}{\gamma \hbar} \td{x}(t') - \frac{8 \lambda}{\beta \gamma^2\hbar^2} \Omega e^{-t'} f_2(t') - \frac{4 \lambda}{\beta \gamma^2 \hbar^2} e^{-t'} f_3(t'),\nonumber\\
&& \ddot{f}_1(t') - e^{t'} \dot{\td{y}}_1(t') = e^{t'}\td{y}_1(t') + \frac{E_{12}}{\gamma} e^{t'}\eta_2 \td{y}_2(t') - \left(\frac{E_{12}}{\gamma}\right)^2 f_1(t'),\nonumber\\
&& \dot{f}_2(t') = e^{t} y_2(t),\nonumber\\
&& \ddot{f}_3(t') - e^{t'} \eta_3 \dot{\td{y}}_2(t') = e^{t'}
\eta_3\td{y}_2(t') -\frac{E_{12}}{\gamma} \eta_2{\gamma}
e^{t'}\td{y}_1(t') - \left(\frac{E_{12}}{\gamma}\right)^2 f_3(t')~
,\label{eq20}
\end{eqnarray}
where overdots denote derivatives with respect to $t^{\prime}$.
These equations  can be efficiently solved numerically.

For comparison with other studies, results are given below for the
particular initial conditions:  $ \rho_{11}(0) = \td{x}(0) =1,
~~\td{y}_1(0) =  \td{y}_2(0)=0,
~~f_1(0)=f_2(0)=f_3(0)=\dot{f}_1(0)\dot{f}_3(0)=0$.
These initial conditions (corresponding to
all the population being on site 1, and no
coherences), are those which have been used extensively in
previous investigations [see Ref.\cite{if}] but are somewhat
unphysical, because they lack initial coherences which become
important in photo-excitation.  We treat this problem of initial
conditions and state preparation with a more plausible model in
Section VII.


\section{Energy representation and Markovian limit}
\subsection{Formalism}
To consider the Markov approximation, we note that it
is particularly simple to invoke in the energy
representation. Hence, below we first utilize the energy basis
and then convert the result back to the site representation
for comparison with the non-Markovian solution.

The Markov approximation can be performed when the time scale on
which the envelope of the density matrix decays is much longer
than the decay time of the phonon correlation function \cite{kuhn}.
Then one can introduce the following approximation:
\begin{equation}
\rho_{ab}^e(t-\tau) \equiv e^{- i \omega_{ab}
(t-\tau)}\td{\rho}_{ab}^e(t-\tau) \simeq e^{- i \omega_{ab}
(t-\tau)}\td{\rho}_{ab}^e(t) = e^{i \omega_{ab} \tau}
\rho_{ab}^e(t). \label{eq21} \end{equation}

As discussed in Ref. \cite{if}, the non-Markovian regime is marked by slow dissipation
of the reorganization energy (i.e., the slow decay of the phonon correlation
function as compared to relaxation dynamics time scale, the decay
of the envelope part of the density matrix). Transitions occur in
accord with the vertical Franck-Condon principle. In the Markovian
regime phonon relaxation is very fast (e.g., large $\gamma$) as
compared to the decay of the envelope of the density matrix.
 Thus, phonons remain
effectively in equilibrium during the EET process in the
Markovian regime \cite{if}.

To obtain the equations in the Markov approximation, Eq. (17) is
first converted to dimensionless form with $\tau' = \gamma \tau$
and $t' = \gamma t$. Putting $t-\tau = \tau'$ in the resulting
equation in the energy representation and then implementing the above
approximation on the density matrix elements allows the time
integration to be performed easily for the case of exponential
phonon correlation function [Eq. (11)]. The result is the set of
Markovian equations;
\begin{eqnarray}
&&\dot{\td{\rho}}_{ab}^e(t') = - i \bar{\omega}_{ab} \td{\rho}_{ab}^e(t') \nonumber\\
&&-\frac{2\lambda}{\beta \hbar^2\gamma^2} \sum_{i,c,d} \left( \frac{V_i^{ac} V_i^{cd}}{1- i \bar{\omega}_{dc}}\td{\rho}_{db}^e(t') - \frac{V_i^{ac} V_i^{db}}{1+ i \bar{\omega}_{db}}\td{\rho}_{cd}^e(t') \right) \nonumber\\
&& + \frac{2\lambda}{\beta \hbar^2\gamma^2} \sum_{i,c,d} \left( \frac{V_i^{ac} V_i^{db}}{1- i \bar{\omega}_{ca}}\td{\rho}_{cd}^e(t') -
 \frac{V_i^{cd} V_i^{db}}{1+ i \bar{\omega}_{cd}}\td{\rho}_{ac}^e(t') \right) \label{eq22}.
\end{eqnarray}
Here $\rho^e_{ab}(t'/\gamma) \equiv \td{\rho}^e_{ab}(t'), ~~
\bar{\omega}_{ab} \equiv \frac{\omega_{ab}}{\gamma}$. Equation
(\ref{eq22}) constitutes a system of coupled ordinary differential
equations that can be solved with given initial conditions.

The results can then be transformed back to the site representation
using the transformation
\begin{equation}
\rho_{ij}(t) = \la i|{\rho}(t)|j\ra = \sum_{a,b}\la i|e_a\ra
\rho_{ab}^e \la e_b|j\ra. \label{eq23}
\end{equation}
where $\rho_{ij}$ is in site representation, $\rho_{ab}^e$ is
 in energy representation, and $i,j,a,b \in \{1,2\}$. Equation (\ref{eq23})
constitutes four linear equations that provides the 
relationship between the representations.

\subsection{Limiting Cases: Analytical Results}
\subsubsection{Strong Coupling Case: $J \gg \Delta$}

For $J \gg \Delta$, we have [from Eq. (\ref{eq14})] $\alpha_1 \simeq 1,~~\alpha_2 \simeq -1,$ and $V_1^{ij} \simeq 1/2$ for $i = j$ and $\simeq -1/2$ for $i \ne j~~(\{i,j\} = 1,2)$ and $V_2^{i,j} \simeq 1/2$ for all $i,j$. One can then analytically solve Eqs. (\ref{eq22}) to obtain the simple expression
\begin{equation}
\tilde{\rho}_{11}^e(t') = \frac{1}{2} (e^{- \frac{4\lambda}{\beta(4 J^2 + \hbar^2 \gamma^2)} t'} + 1),
\end{equation}
for the traditional initial conditions $\tilde{\rho}_{11}^e(t'=0) =1,~~ \tilde{\rho}_{12}^e(t'=0) = \tilde{\rho}_{21}^e(t'=0) = 0$. The Markov approximation can be performed when the time scale on which the envelope
of the density matrix decays is much longer than the decay time of the phonon auto-correlation function. Hence, $ \frac{4\lambda}{\beta(4 J^2 + \hbar^2 \gamma^2)} \ll 1$ must hold for the Markov approximation to be valid in the $J \gg \Delta$ domain. (Note that the decay time constant for the phonon auto-correlation function is unity, since we defined $t' = \gamma t$.)

\subsubsection{Weak Coupling Case: $J \ll \Delta$}
For $J \ll \Delta$, we have [from Eq. (\ref{eq14})], $\alpha_{1,2}= \frac{1}{2J}(\Delta \mp \sqrt{\Delta^2 + 4J^2}) \simeq \frac{\Delta}{2 J}(1\mp1)$. Hence, in this domain $\alpha_1 \simeq 0$, and $\alpha_2 \simeq \Delta/J$. This leads to $V_1^{11} =V_1^{12}=V_1^{21} \simeq 0,~~V_1^{22}\simeq 1$. $V_2^{11} \simeq 1,~~V_2^{12}=V_2^{21}\simeq J/\Delta$, and $V_2^{22} = (\frac{J}{\Delta})^2$. From Eq. (\ref{eq22}) we then have

\begin{eqnarray}
&&\dot{\tilde{\rho}}_{11}^e(t')= \nonumber\\
&&\frac{2/\lambda}{\hbar^2\beta \gamma^2}\left(2(J/\Delta)^2 (\Gamma +\Gamma^\ast) - 4(J/\Delta)^2 (\Gamma +\Gamma^\ast) \tilde{\rho}_{11}^e(t') + 2 (J/\Delta) (\tilde{\rho}_{12}^e(t') + \tilde{\rho}_{21}^e(t')) \right), \label{eq42}
\end{eqnarray}

\begin{eqnarray}
&& \dot{\tilde{\rho}}_{12}^e(t') = \frac{ i \Delta}{\hbar \gamma}\tilde{\rho}_{12}^e(t')+\nonumber\\
&& + \frac{4\lambda}{\hbar^2\beta \gamma^2}\left((J/\Delta)\Gamma^\ast (2 \tilde{\rho}_{11}^e(t') -1) - (1 + 2 \Gamma (J/\Delta)^2) \tilde{\rho}_{12}^e(t') + 2 (J/\Delta)^2 \Gamma^\ast \tilde{\rho}_{21}^e(t')\right). \label{eq43}
\end{eqnarray}
where $\Gamma = \frac{1}{1+ i\frac{\Delta}{\hbar\gamma}}$.

The reduction from a large number (33) of terms in Eq. (\ref{eq22}) to the smaller number of terms [in Eq. (\ref{eq43})] is made possible by neglecting small terms of the order of $(J/\Delta)^3$. However, even in this approximation the above equations do not admit a simple analytic solution, and no simple analytic expression can be given for the range of validity of the Markov approximation. Hence, we invoke a further approximation, neglecting second order terms $(J/\Delta)^2$ as compared to first order $J/\Delta$. By separating real and imaginary parts as $\tilde{\rho}_{12}^e(t') = x(t') + i y(t')$ and writing $\tilde{\rho}_{11}^e(t') = r(t') $, Eqs. (\ref{eq42}) and (\ref{eq43}) become

\begin{eqnarray}
&&\dot{r}(t') = (\frac{J}{\Delta})\frac{4\lambda}{\hbar^2\beta \gamma^2} x(t'),\nonumber\\
&&\dot{x}(t') = - \frac{\Delta}{\hbar\gamma}y(t') - \frac{4\lambda}{\hbar^2\beta \gamma^2} x(t'),\nonumber\\
&&\dot{y}(t') = \frac{\Delta}{\hbar\gamma}x(t') - \frac{4\lambda}{\hbar^2\beta \gamma^2} y(t'). \label{eq44}
\end{eqnarray}

These coupled ordinary differential equations have the straightforward solution:

\begin{eqnarray}
&&r(t') = \frac{1}{\eta^2 + \xi^2} (\eta^2 +\xi^2 +[a\xi-b\eta]\ep \xi \nonumber\\
&& [b\eta-a\xi] \ep\xi \cos(\eta t')e^{-\xi t'} + [a\eta + b\xi] \ep\xi\sin(\eta t')e^{-\xi t'}),\nonumber\\
&& x(t') = e^{-\xi t'} (a \cos(\eta t')-b\sin(\eta t')),\nonumber\\
&&y(t') = e^{-\xi t'} (a \sin(\eta t') + b \cos(\eta t')). \label{eq45}
\end{eqnarray}
with initial conditions $r(t'=0)=1,~~x(t'=0)=a,~~y(t'=0)=b$. Here $\xi = \frac{4\lambda}{\hbar^2 \beta \gamma^2},~~\eta = \frac{\Delta}{\hbar\gamma},$ and $\ep = \frac{J}{\Delta}$. Interestingly, for $a=b=0$ the density matrix elements do not change with time. This is due to our approximations of only retaining terms first order in $J/\Delta$. Thus, for the condition of Markov approximation to hold requires $ \xi= \frac{4\lambda}{\beta \hbar^2 \gamma^2} \ll 1$ (again noting that the decay time constant for the phonon auto-correlation function is unity since $t' = \gamma t$).

\begin{figure}[htbp]
\centering
\begin{tabular}{cc}
\includegraphics[height = 5cm, width = 7cm]{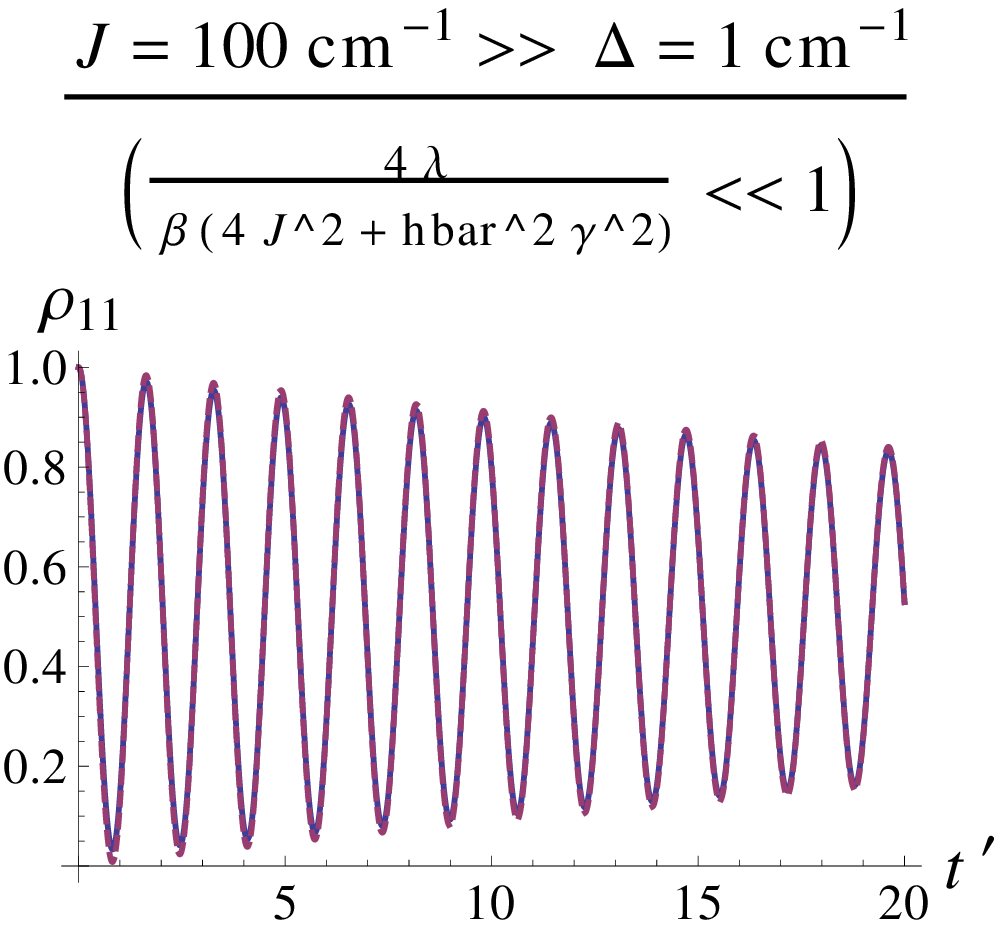}&
\includegraphics[height = 5cm, width = 7cm]{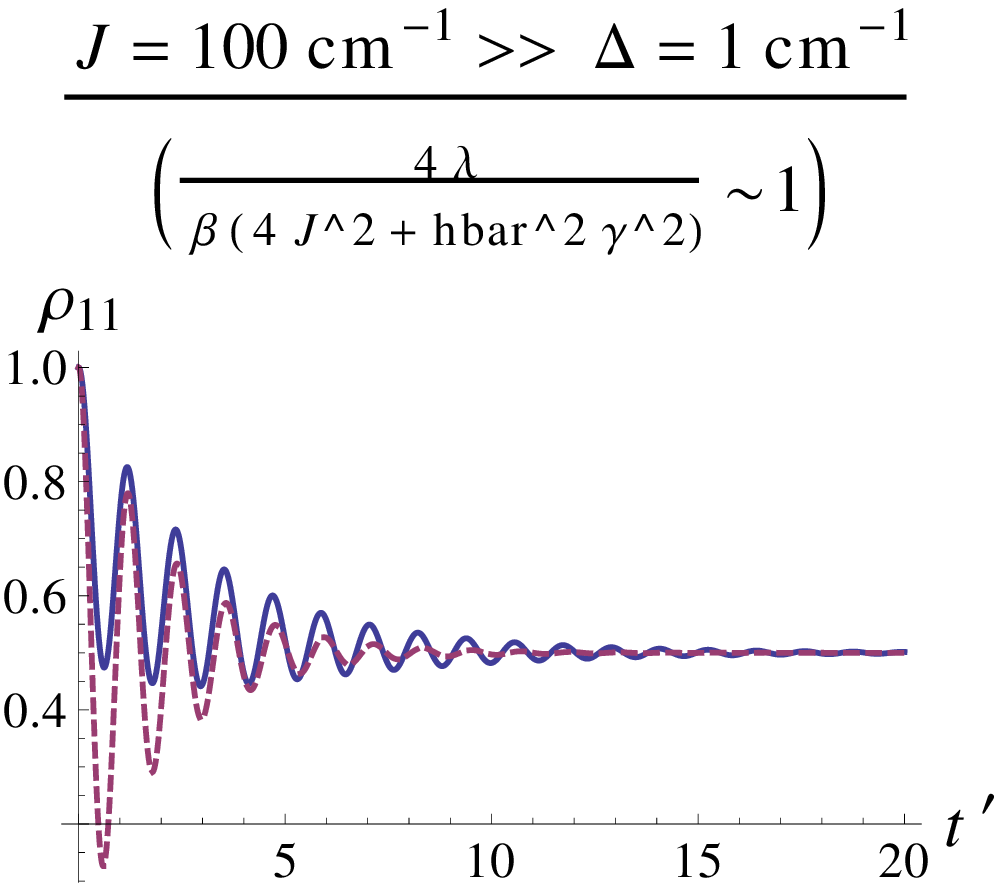}\\
\includegraphics[height = 5cm, width = 7cm]{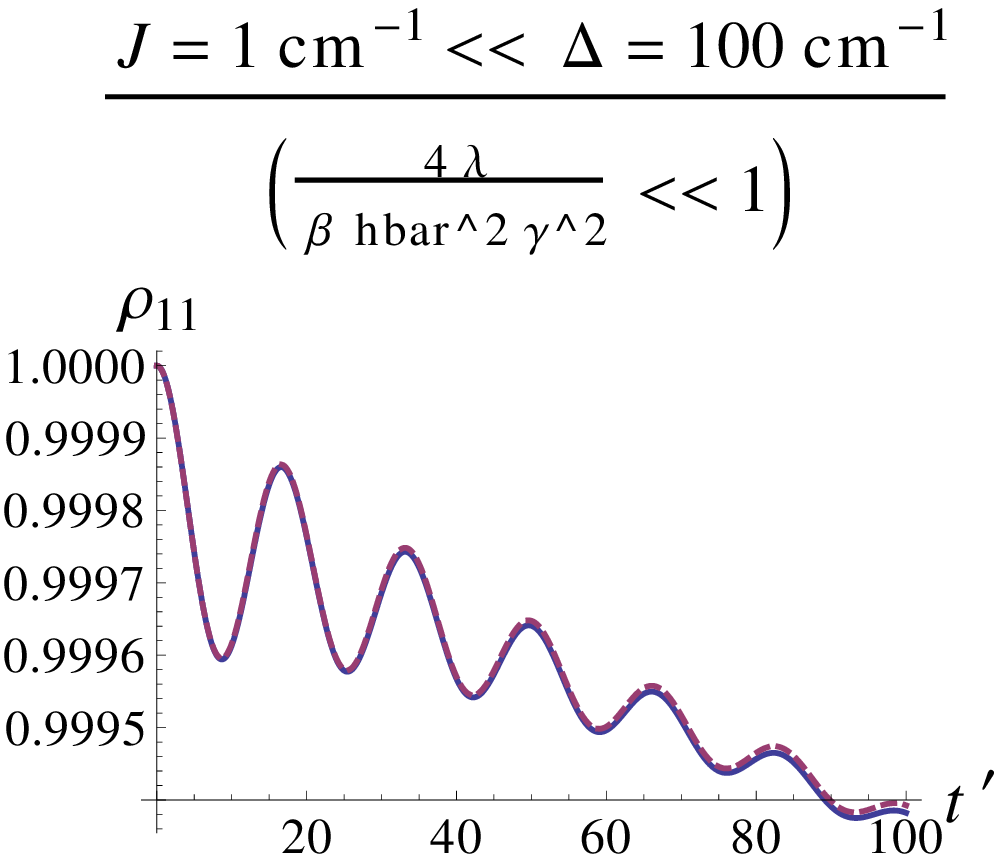}&
\includegraphics[height = 5cm, width = 7cm]{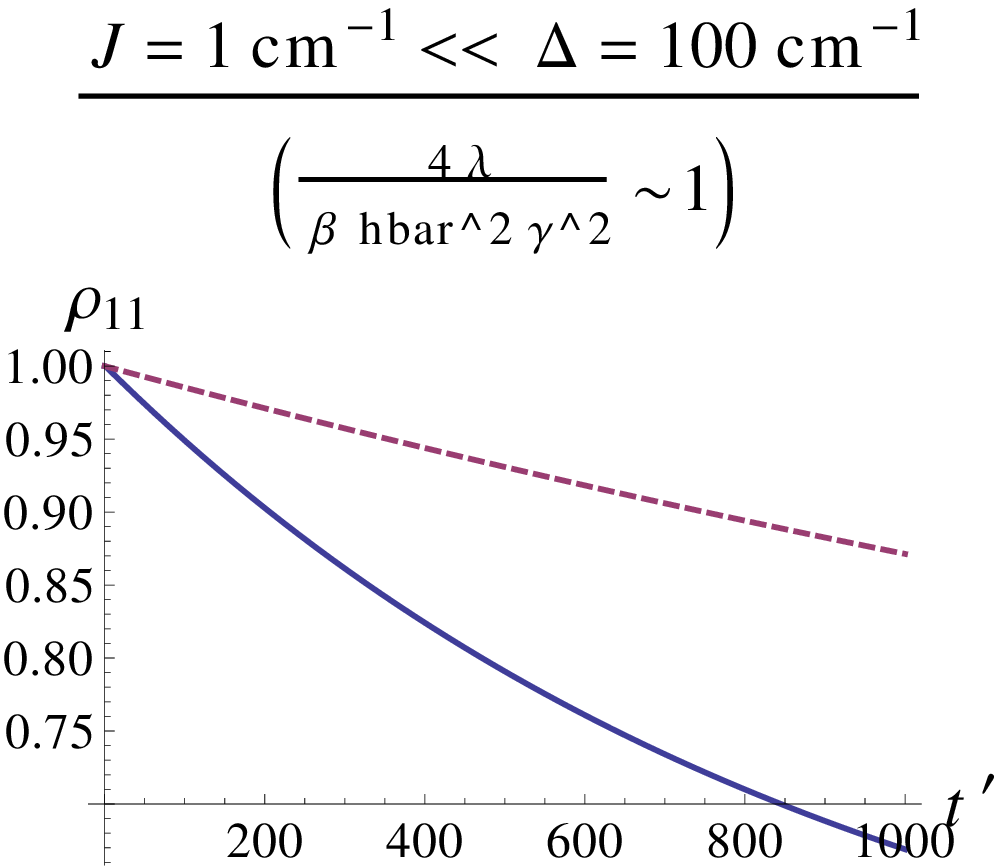}\\
\end{tabular}
\caption{Verification of the analytic inequalities (Table I). Time evolution of population on site 1: blue (solid) curve is the non-Markovian solution and red (dotted) curve is the
Markovian approximation. Upper row (case $J>>\Delta$): $\frac{4\lambda}{\beta (4 J^2 + \hbar^2 \gamma^2)}=0.04<<1, ~~\lambda =2$ (upper left graph), and $\frac{4\lambda}{\beta (4 J^2 + \hbar^2 \gamma^2)}=0.9 \sim 1,~~\lambda = 50$(upper right graph). Lower row (case $J<<\Delta$): 
$\frac{4\lambda}{\beta \hbar^2 \gamma^2}=0.02<<1, ~~\lambda =2$ (lower left graph), and $\frac{4\lambda}{\beta \hbar^2 \gamma^2} \sim 3,~~\lambda = 10$ (lower right graph). 
Clearly, graphs are in accord with Table I. Time $ t'$ is the dimensionless time,
$10$ units on this scale are equivalent to one ps.} \label{fig1new}
\end{figure}

 These analytic results are summarized in Table I, and 
these inequalities have been numerically verified (e.g., see 
Fig. \ref{fig1new}). Note that the $J \gg \Delta$ result goes over to the $J \ll \Delta$ result as $J$ gets smaller.

 \begin{table}
\caption{Regimes of validity of the Markov approximation}
\begin{ruledtabular}
\begin{tabular}{lcr}

Case & Approx. matrix elements & Markovian approximation \\\hline

$J>>\Delta$ & $V_k^{ij} \simeq (-1)^{k(i+j)}\frac{1}{2},~~i,j,k=1,2 $ & $\frac{4\lambda}{\beta (4 J^2 + \hbar^2 \gamma^2)}<<1 $\\\hline

$J<< \Delta$  & $~V_k^{ij}\simeq \delta_{k1}\delta_{i2}\delta_{j2} +\delta_{k2} (\delta_{i1}\delta_{j1} + (\frac{J}{\Delta})^2\delta_{i2}\delta_{j2} + \frac{J}{\Delta}(\delta_{i1}\delta_{j2}+\delta_{i2}\delta_{j1}))$  & $\frac{4 \lambda}{\hbar^2 \beta \gamma^2}<<1$ \\

\end{tabular}
\end{ruledtabular}
\end{table}

\subsection{Computational Results}

\begin{figure}
\centering
\begin{tabular}{ccc}
\includegraphics[height = 4.5cm, width = 5.5cm]{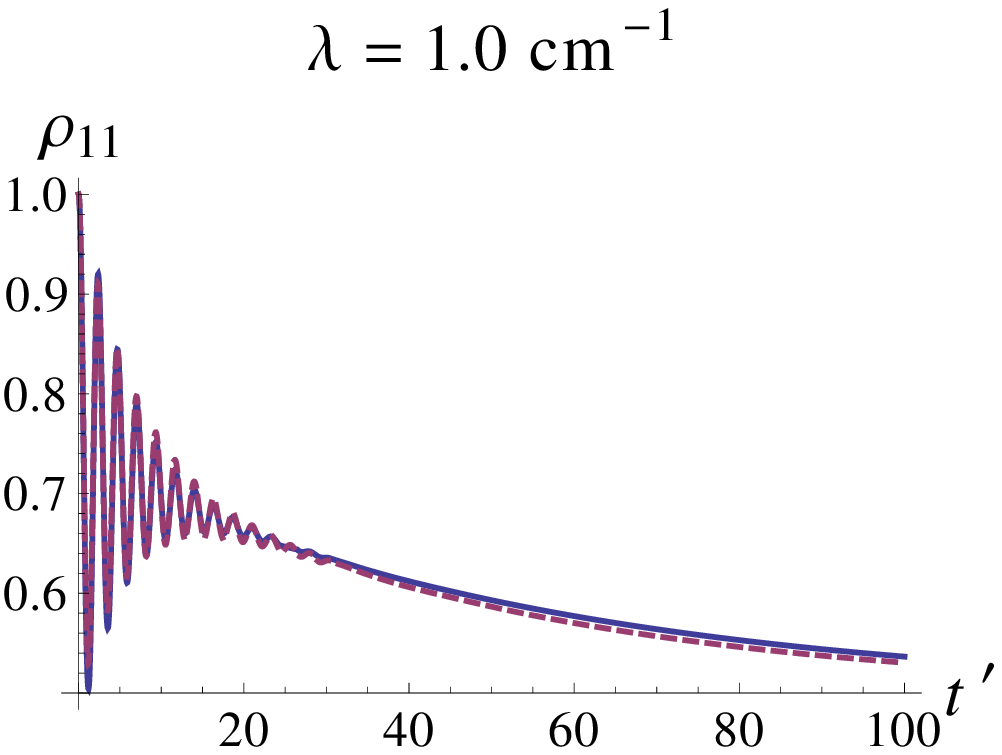}&
\includegraphics[height = 4.5cm, width = 5.5cm]{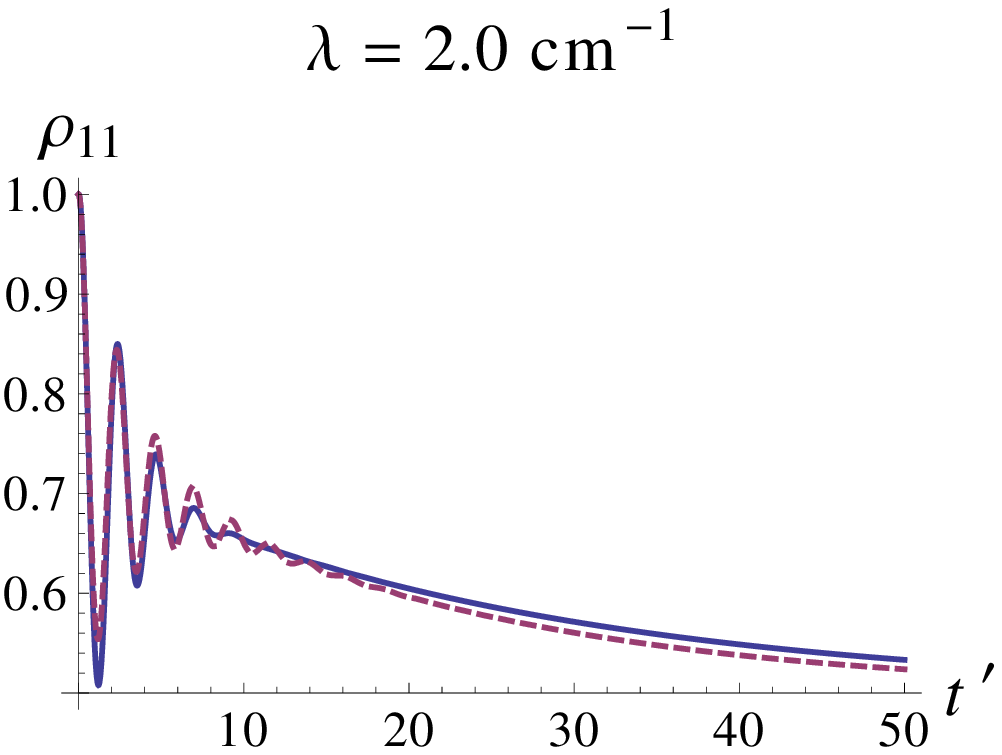}&
\includegraphics[height = 4.5cm, width = 5.5cm]{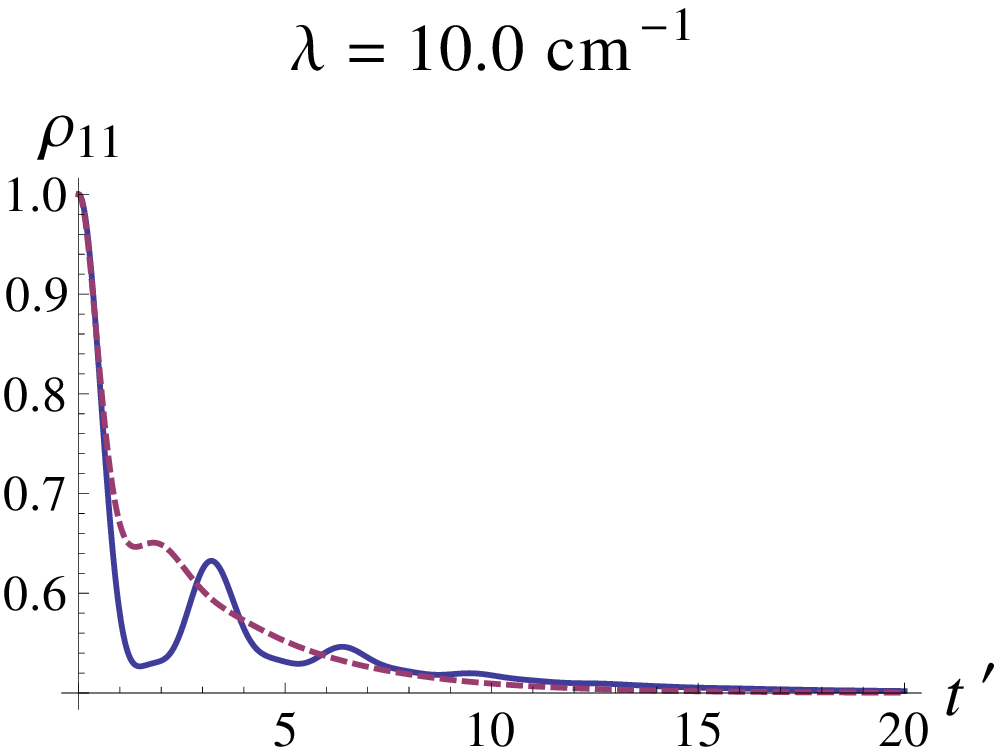}\\
\includegraphics[height = 4.5cm, width = 5.5cm]{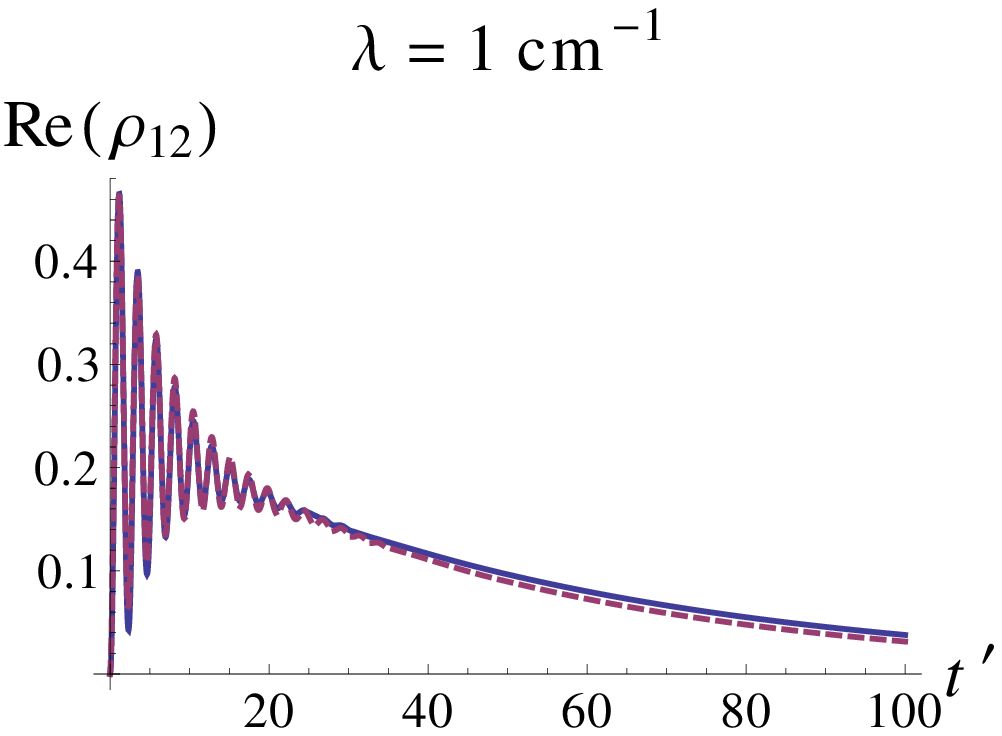}&
\includegraphics[height = 4.5cm, width = 5.5cm]{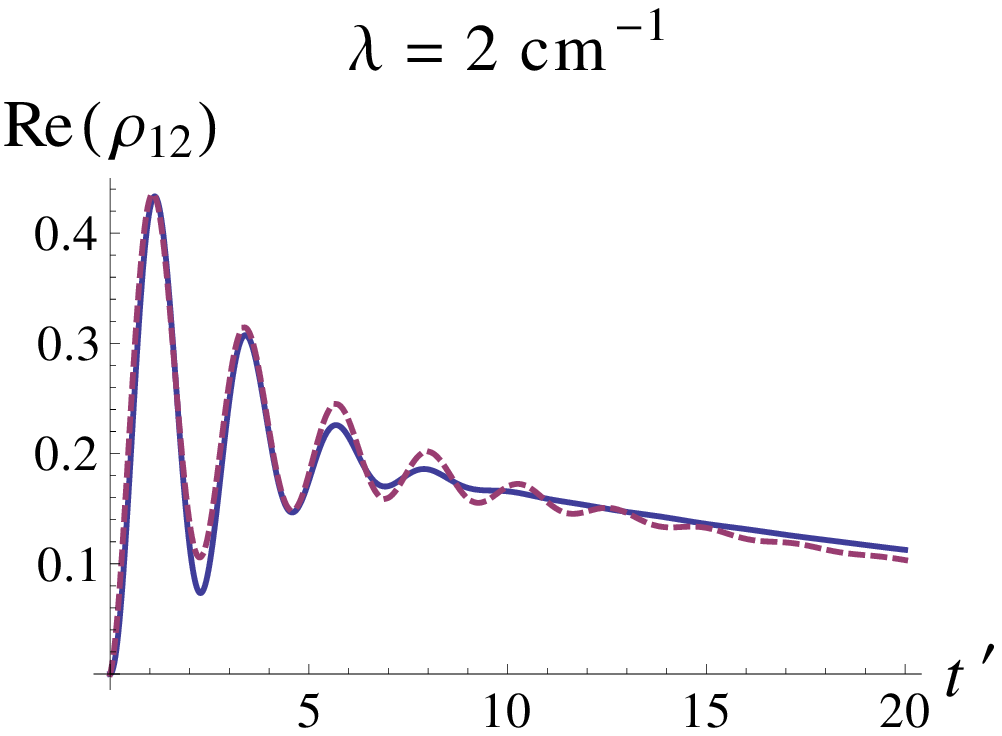}&
\includegraphics[height = 4.5cm, width = 5.5cm]{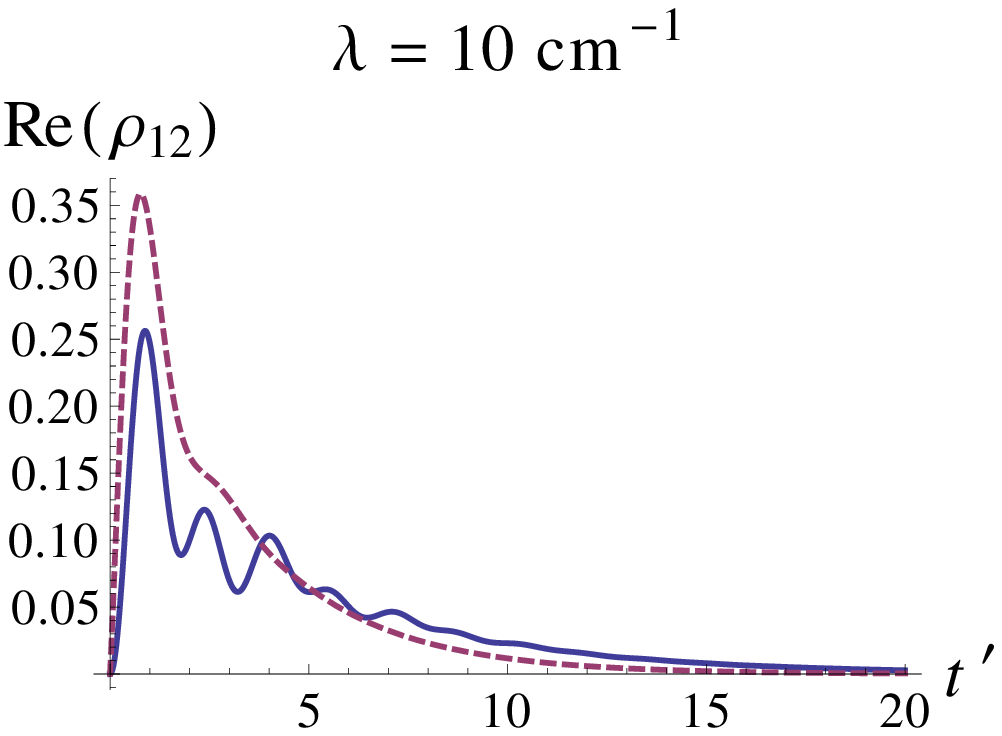}\\
\includegraphics[height = 4.5cm, width = 5.5cm]{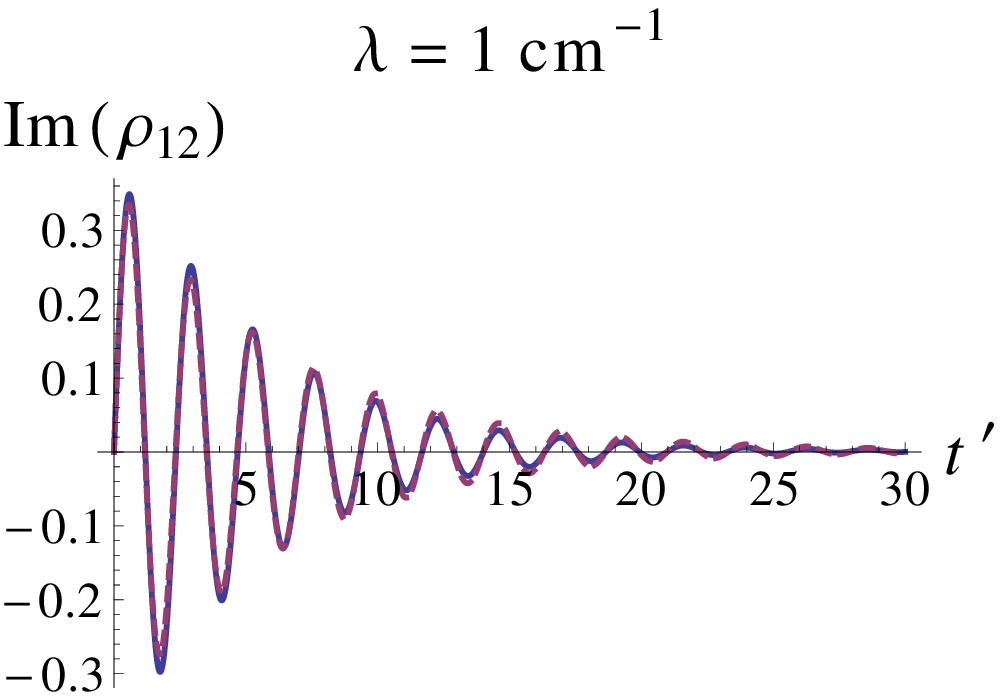}&
\includegraphics[height = 4.5cm, width = 5.5cm]{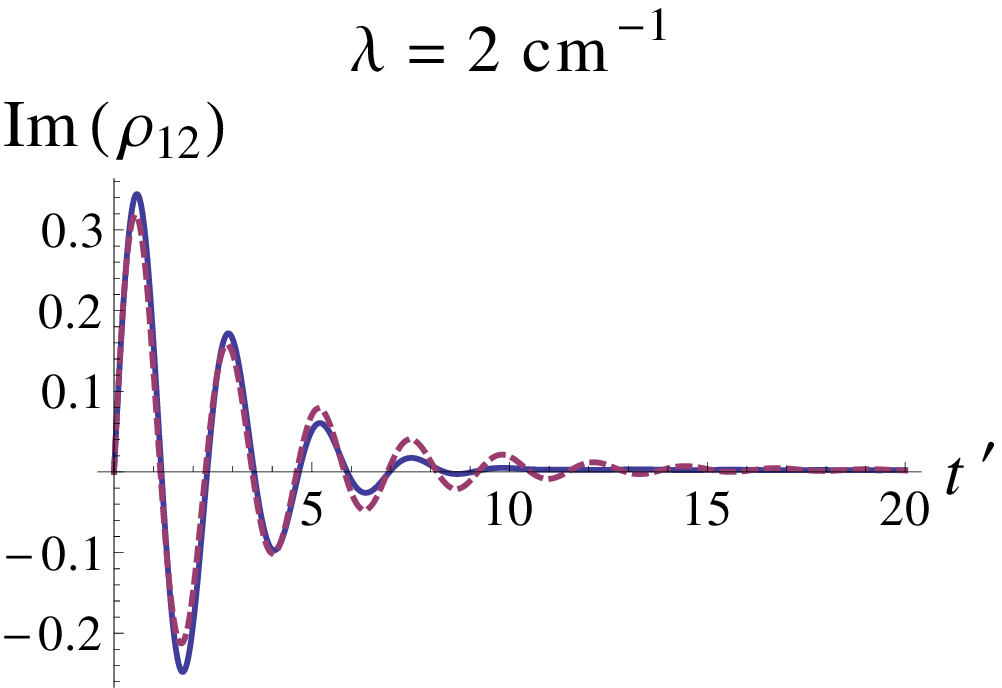}&
\includegraphics[height = 4.5cm, width = 5.5cm]{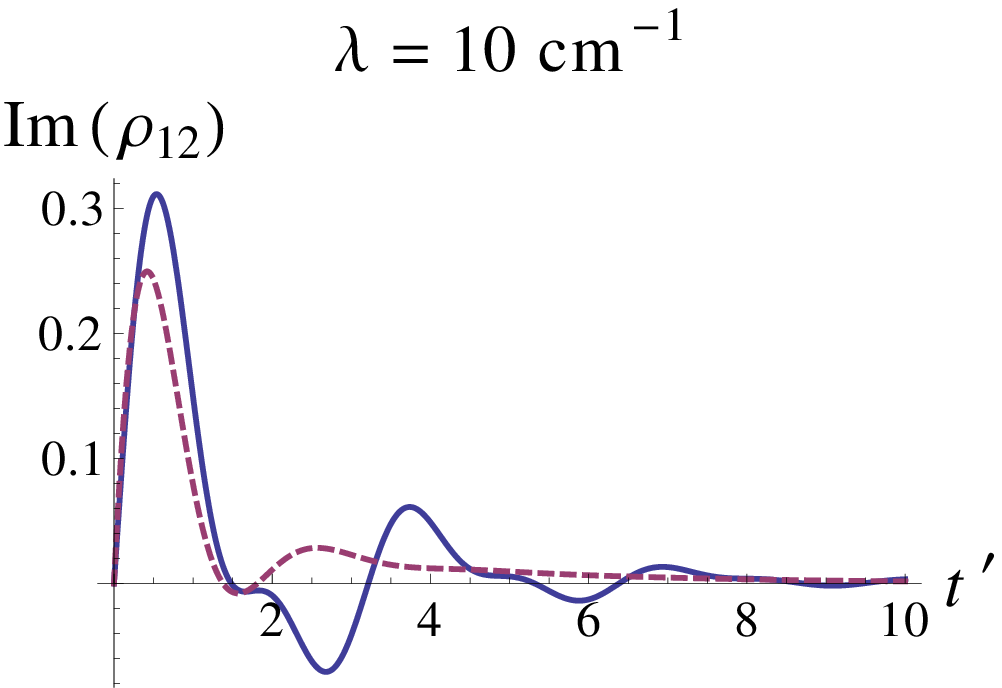}
\end{tabular}
\caption{Time evolution of population on site 1 [$\rho_{11}(t)$]
and the coherences [$\rho_{12}(t)$] [blue (solid) curve is the
non-Markovian solution and red (dotted) curve is the Markovian
approximation], for various values of
$\lambda$ (in cm$^{-1}$). Other parameter are: $\Delta = 100 cm^{-1},~~J = 50 cm^{-1},~~\gamma = 10^{13} sec^{-1}$.
The breakdown of the
Markov approximation at $\lambda = 10 ~$cm$^{-1}$ is clearly visible. The time $ t'$ is dimensionless, with
10 units on this scale being equivalent to one ps.}\label{fig1}
\end{figure}

In other parameter regimes, the validity of the Markov
approximation  [Eq.  (\ref{eq22})] must be determined by
numerical  comparisons  with  the  exact  result  [Eq.
(\ref{eq20})].
Figure \ref{fig1} compares the solution for the Markovian master equation
to the non-Markovian results for the
standard electronic coupling
parameter values in photosynthetic EET:
$\gamma^{-1} = 100~\textrm{fs},~~J = 50 ~\textrm{cm}^{-1},~~
\Delta = 100~\textrm{cm}^{-1},~~T =300~$K , a regime in which the estimates in
Table I do not apply. The initial excitation is
assumed to be on site one.
The Markovian approximation is seen to be very good
for $\lambda = 1 ~ \textrm{cm}^{-1}$,
fair for $\lambda = 2 ~ \textrm{cm}^{-1}$ and
invalid for reorganization
energies $\lambda \geq 10 ~ \textrm{cm}^{-1}$.

\begin{figure}
\centering
\begin{tabular}{ccc}
\includegraphics[height = 4cm, width = 5cm]{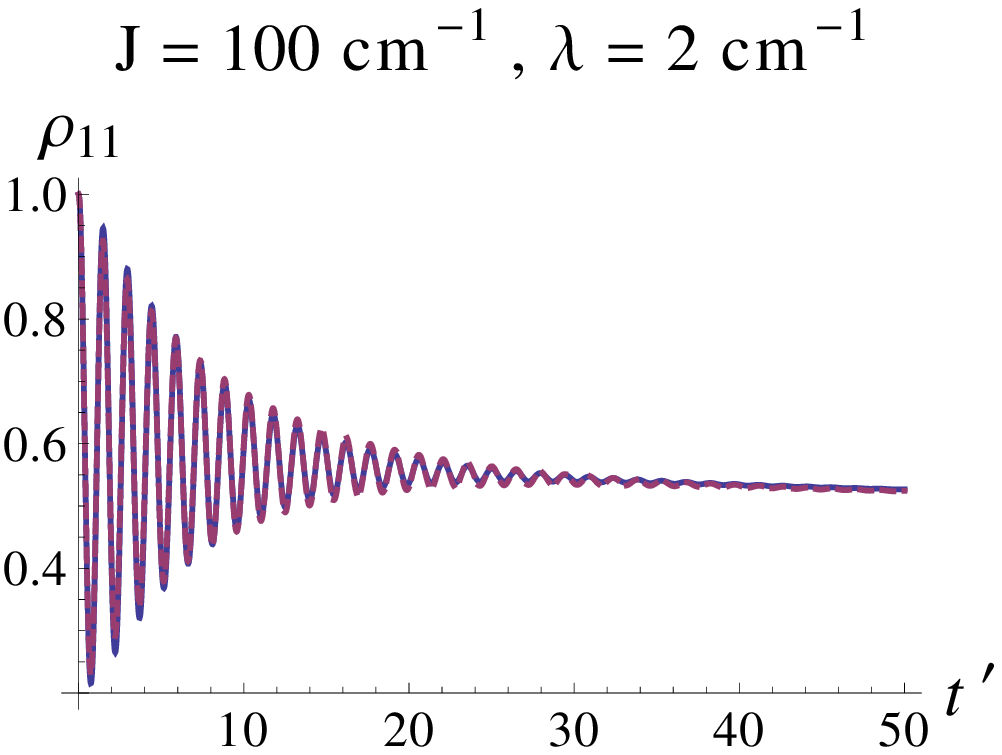}&
\includegraphics[height = 4cm, width = 5cm]{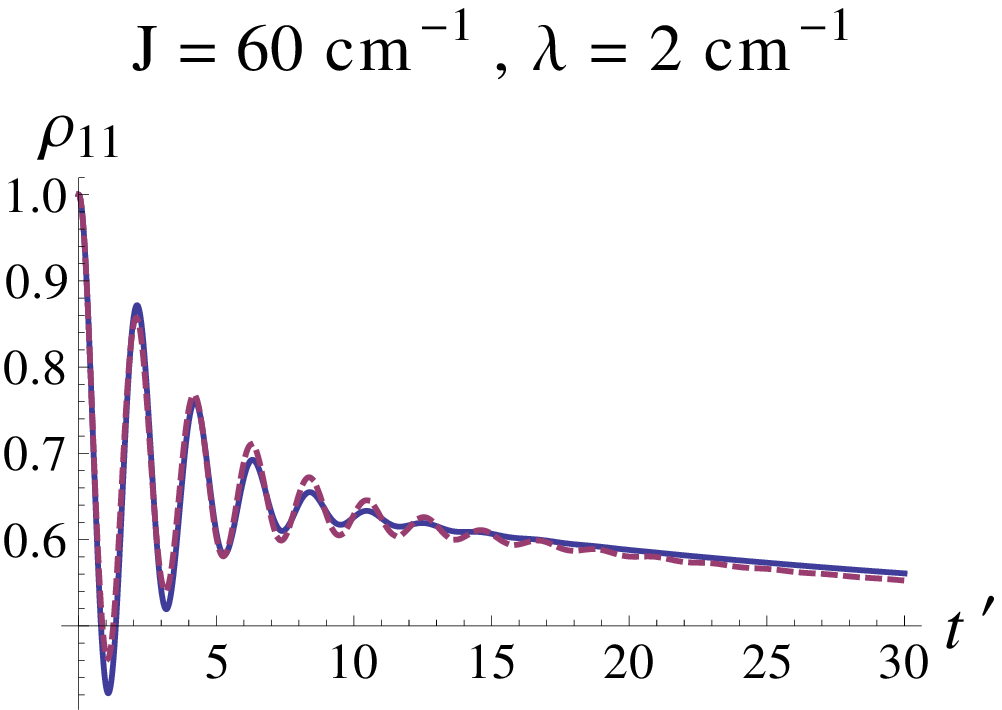}&
\includegraphics[height = 4cm, width = 5cm]{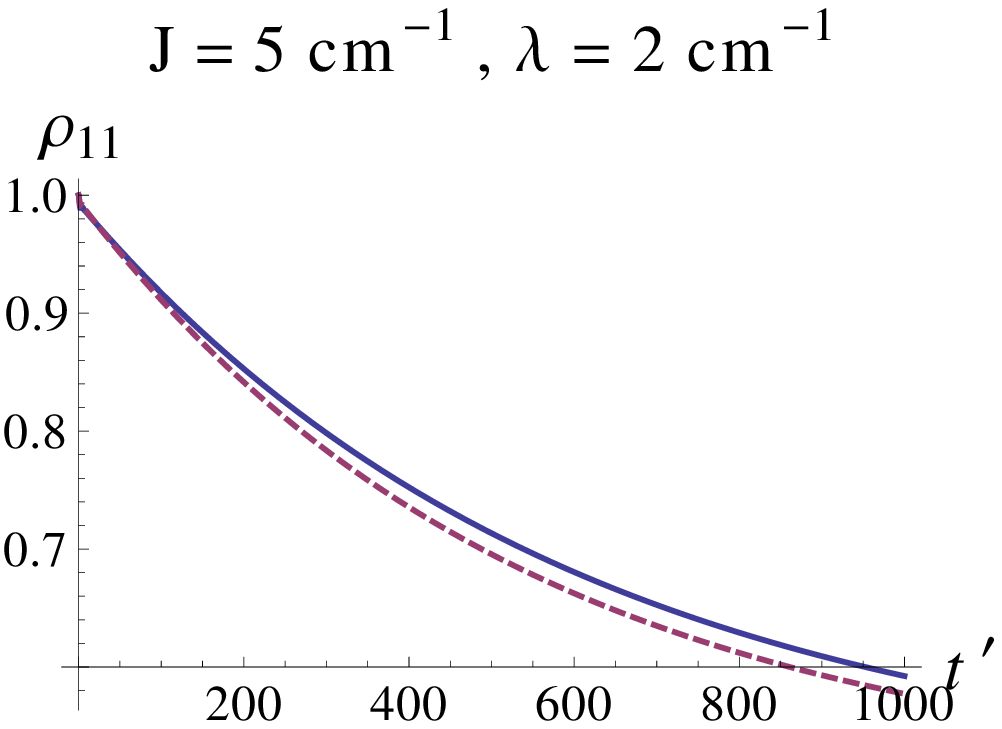}\\
\includegraphics[height = 4cm, width = 5cm]{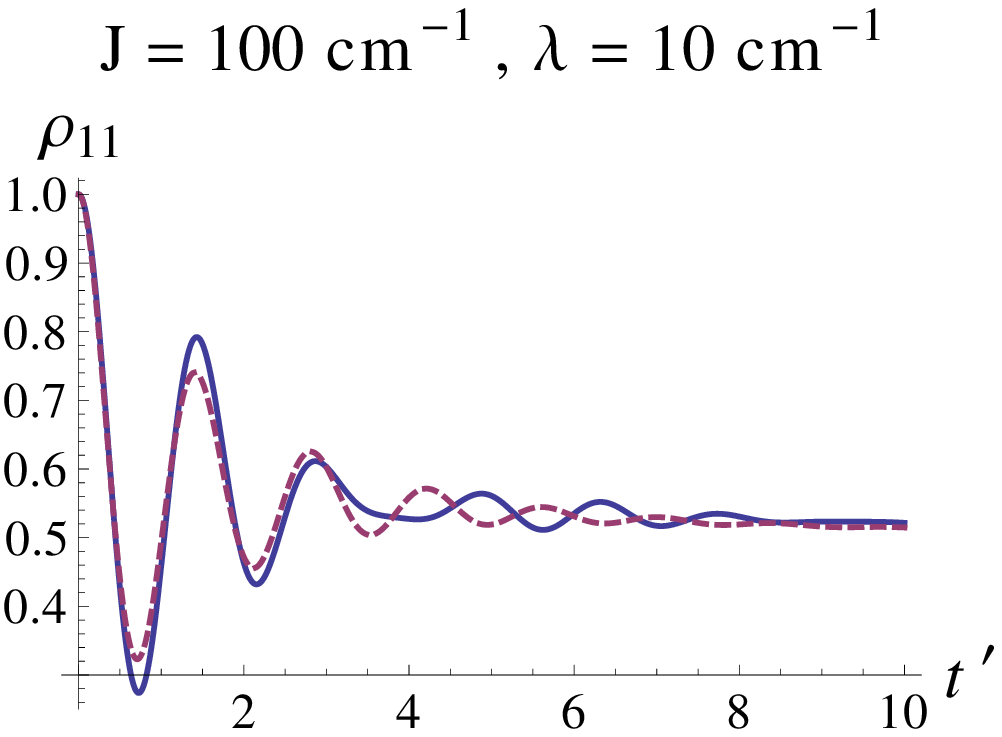}&
\includegraphics[height = 4cm, width = 5cm]{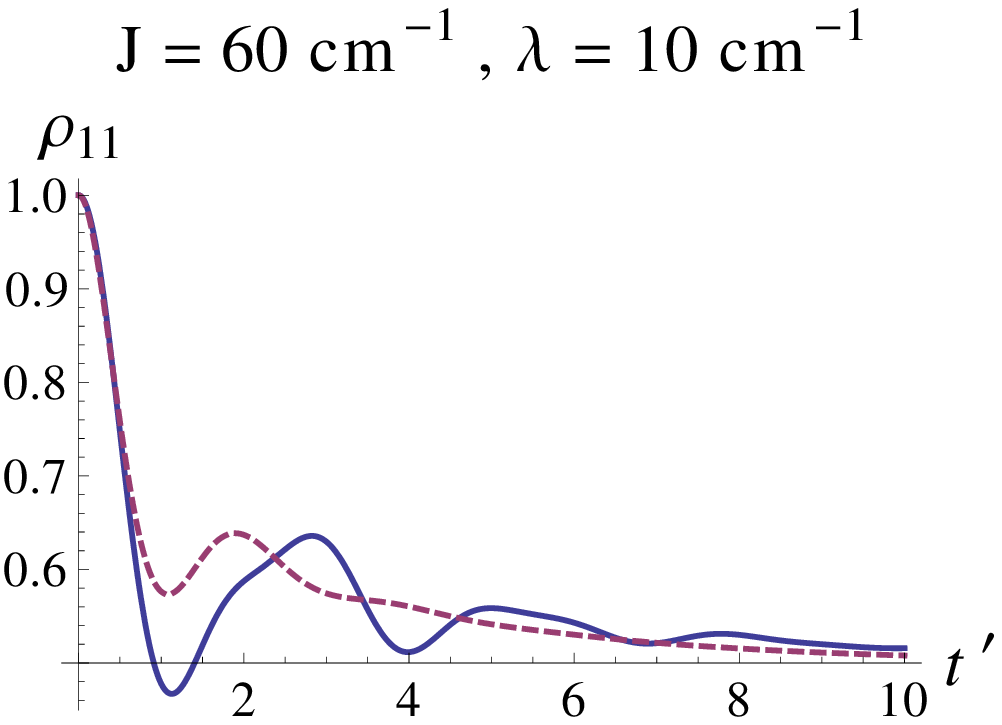}&
\includegraphics[height = 4cm, width = 5cm]{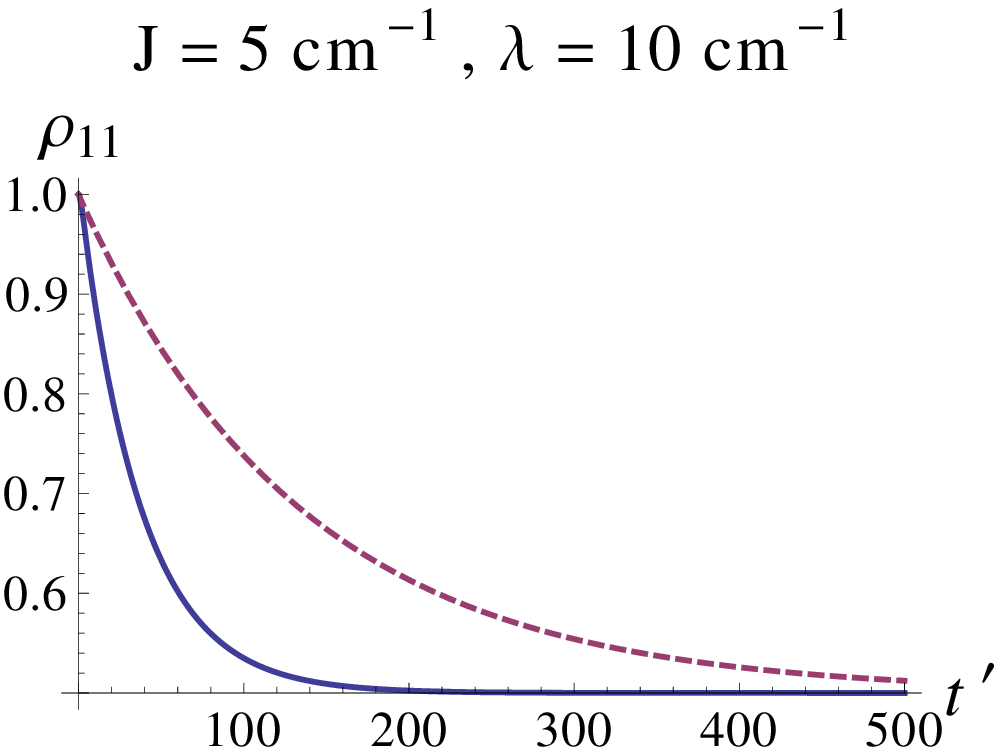}
\end{tabular}
\caption{Time evolution of population on site 1: blue (solid)
curve is the non-Markovian solution and red (dotted) curve is the
Markovian approximation) for various values of the reorganization
energy $\lambda$ and inter-site coupling $J$. The level separation
$\Delta = 100~\textrm{cm}^{-1}$ and $\gamma^{-1} = 100$ fs. It
is clear that Markovian approximation is poor for large
$\lambda$ and small $J$. Time $ t'$ is the dimensionless time,
$10$ units on this scale are equivalent to one ps.} \label{fig2}
\end{figure}

\begin{figure}
\centering
\begin{tabular}{ccc}
\includegraphics[height = 4cm, width = 5cm]{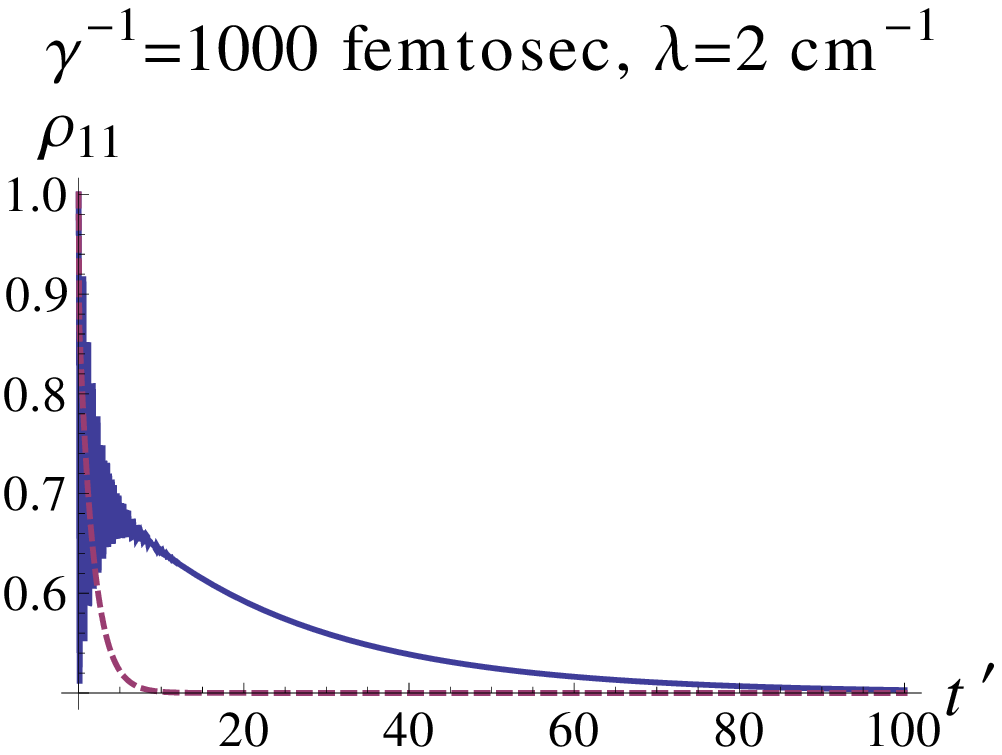}&
\includegraphics[height = 4cm, width = 5cm]{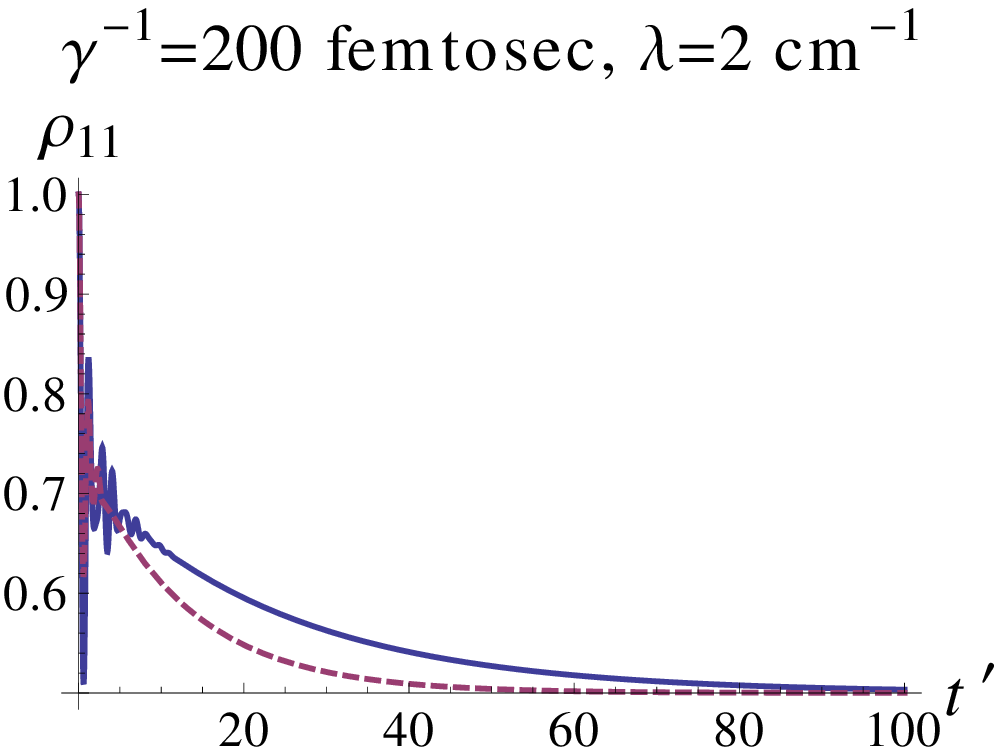}&
\includegraphics[height = 4cm, width = 5cm]{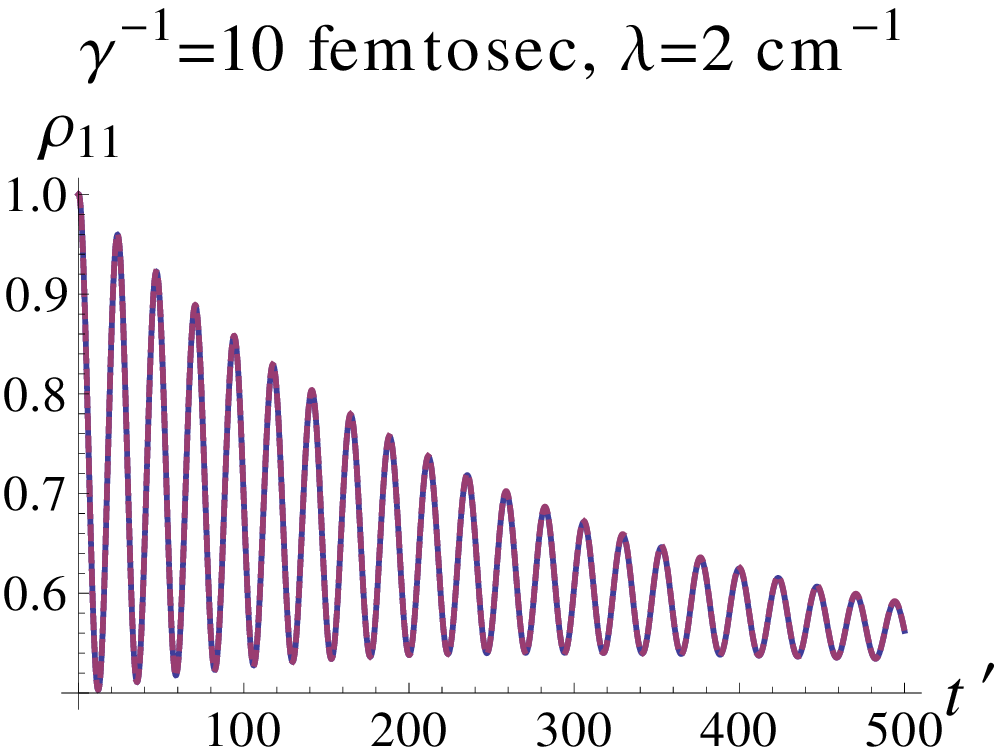}\\
\includegraphics[height = 4cm, width = 5cm]{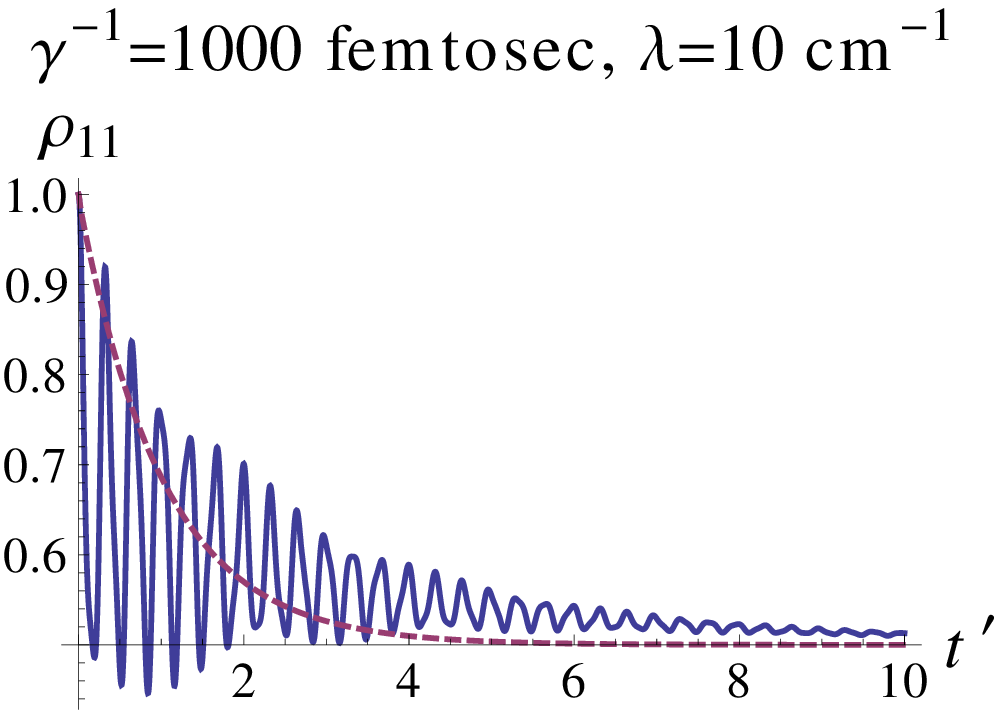}&
\includegraphics[height = 4cm, width = 5cm]{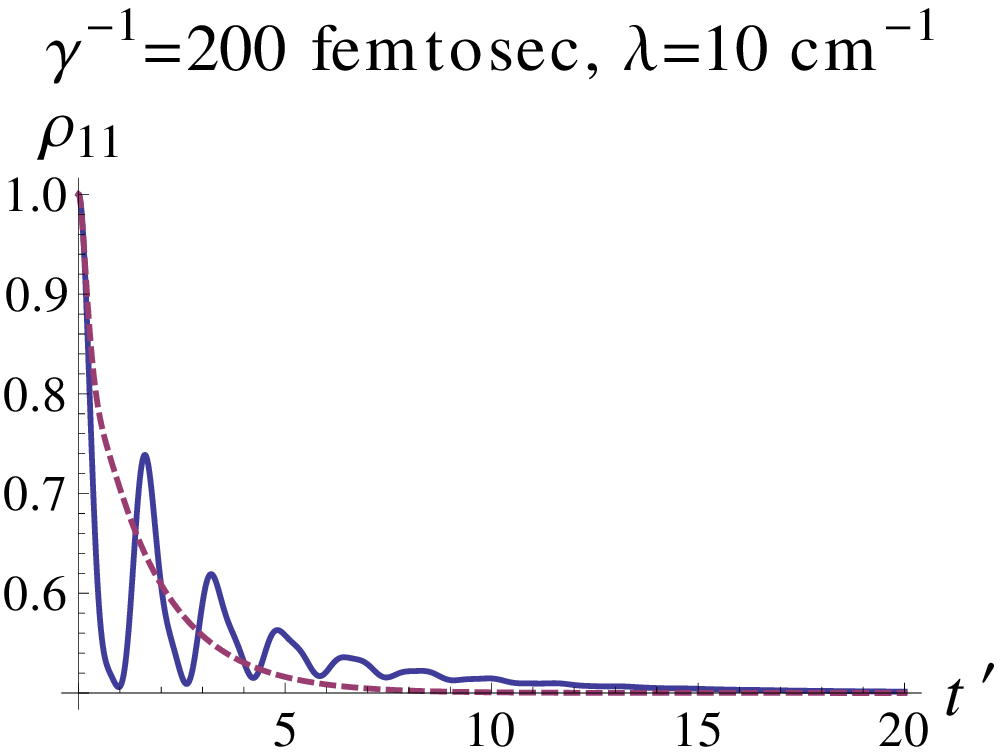}&
\includegraphics[height = 4cm, width = 5cm]{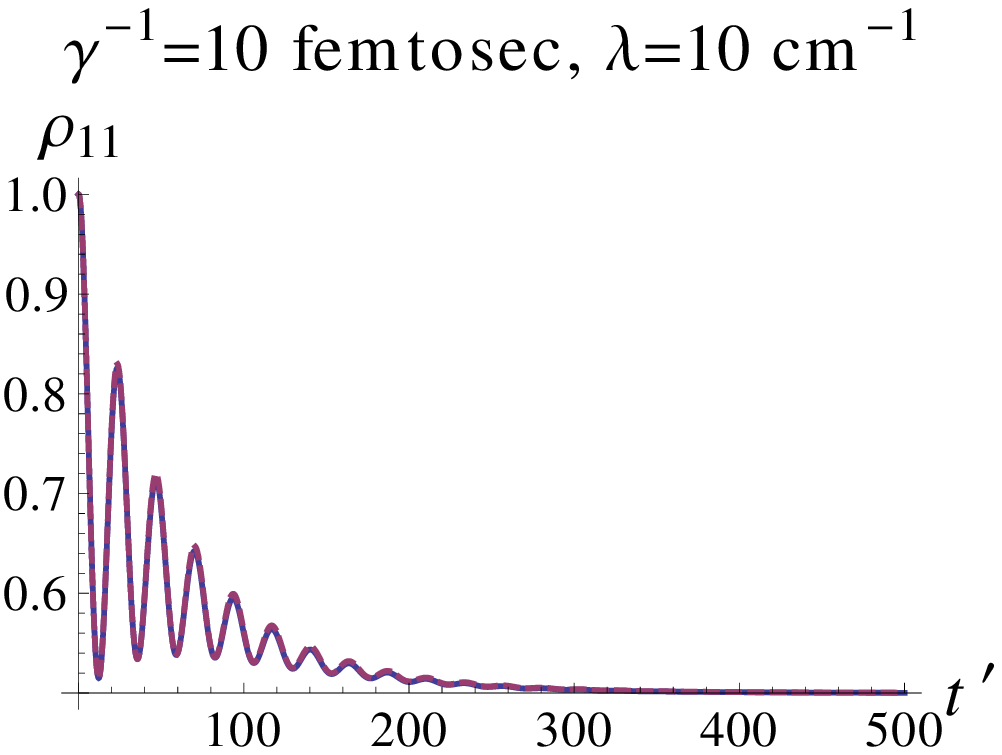}
\end{tabular}
\caption{Time evolution of population on site 1 [blue (solid)
curve is the non-Markovian solution and red (dotted) curve is the
Markovian approximation] for various values of the reorganization
energy $\lambda$ and phonon relaxation time $\gamma^{-1}$. The
level separation $\Delta = 100~\textrm{cm}^{-1}$ and $J =
50~\textrm{cm}^{-1}$. It is clear that Markovian approximation is
poor for large relaxation times $\gamma^{-1}$ and small
$\lambda$, and it better for small $\gamma^{-1}$. $t'$ is the scaled time, as in the figure above. }\label{fig3}
\end{figure}

To explore regimes of validity of the Markovian approximation for
other values of the physical constants, we present
sample results in Figs. \ref{fig2} and \ref{fig3}, obtained by varying $(J,\lambda)$ and
$(\gamma^{-1}, \lambda)$, keeping $\Delta=100$ cm$^{-1}$. The results show 
for these cases that the
Markovian approximation is poor for large $\lambda$ and
``small' $J$ and for large $\gamma^{-1}$ and small $\lambda$.
Other parameter values can be readily examined computationally using this
approach.


\section{Regime of validity of the second-order approximation}

We have considered the master equation up to the second order in
system-bath interaction. The aim of this section is to determine
the system-bath interaction energy range (as represented by the
reorganization energy $\lambda$) over which the
second order master equation [or coupled system of equations (Eq.
(20)] can be used. To do so we compare estimates of the second order
and the fourth order terms. This can be done analytically for the parameter 
regime where $J \sim \Delta$. To do so
we note that up to the second order, with the
master equation written in dimensionless time form [Eq. (20)],
the magnitude of the second order term is of the order of
$\frac{\lambda}{\beta \hbar^2 \gamma^2}$ for the case $J \sim \Delta$. This arises by noting that
$|H_{SR}|^2 \sim
\lambda$, and the integral $\int_0^{t'} d\tau' e^{\tau'-t'} \left[
\cos[\frac{E_{12}}{\gamma}(t'-\tau')] \td{y}_1(\tau') + \eta_2
\sin[\frac{E_{12}}{\gamma}(t'-\tau')]\td{y}_2(\tau') \right] \sim
1$, for the standard set of parameters ($\gamma^{-1} = 100~\textrm{fs},~~J = 20 ~\textrm{cm}^{-1},~~\Delta = 100~\textrm{cm}^{-1},~~T = 300~$K).

Similarly, we can estimate the parameter dependence of the fourth
order term. To estimate this we recall the Nakajima-Zwanzig master
equation (valid to all orders)
\begin{equation}
\frac{\pr \hat{\rho}^I(t)}{\pr t} = -\int_0^t d\tau tr_{R}
\left(\ml_{SR}^I \ms(t,\tau) \mq \ml_{SR}^I(\tau) \hat{R}_{eq}
\right) \hat{\rho}^I(\tau).
\end{equation}
where the time evolution operator is
\begin{eqnarray}
\ms(t,\tau)& \equiv & \mathcal{T}^{\rightarrow} \exp[- i
\int_\tau^t d\tau' \mq \ml_{SR}^I(\tau')]\nonumber \\& = &1- i
\int_\tau^t d\tau' \mq \ml_{SR}^I(\tau') - \int_\tau^t
d\tau_2\int_\tau^{\tau_2} d\tau_1
\mq\ml_{SR}^I(\tau_2)\mq\ml_{SR}^I(\tau_1) + ...
\end{eqnarray}
Here, $\mq = I - \mathcal{P}$ is the well know projection operator and $\ml_{SR}^I$ is the system-bath Liouvillian ($-\frac{i}{\hbar}[H_{SR}^I,.] $). 
The time ordering operator $\mathcal{T}^{\rightarrow}$ orders
time dependent operators from left to right with decreasing time
arguments, to take into account the non-commutation of operators
at different times.

The zeroth order approximation to the time evolution operator $\ms(t,\tau)$
gives the second order quantum master equation, the first
order approximation to the time evolution operator gives the third
order contribution which vanishes as the bath average of odd bath
operators vanish (see Ref. \cite{hpb}), and the second order
approximation to $\mathcal{S}$ gives the fourth order
contribution. In order to estimate the magnitude of the latter
term we write the fourth order term A$_4$ from Nakajima-Zwanzig
equation as
\begin{equation}
\textrm{A}_4 =\frac{1}{\hbar^4} \int_0^t d\tau \int_\tau^t
d\tau_2\int_\tau^{\tau_2} d\tau_1 tr_{R} \left\{ [
H_{SR}^I(t),\mq[ H_{SR}^I(\tau_2),\mq [
H_{SR}^I(\tau_1),\mq[H_{SR}^I(\tau),R_{eq} \hat{\rho}(\tau)]]]]
\right\}.
\end{equation}

We start from the interior commutator $(1-
\mathcal{P})[H_{SR}^I(\tau),R_{eq} \hat{\rho}(\tau)]$ and recall
that $\mathcal{P} \hat{O} = R_{eq} tr_{R}{\hat{O}}$ and $H_{SR}^I
= V_i^I u_i^I$ (where the summation convection is used). The bath
average of single bath operators vanish [$tr_{R} (u_{j}(\tau)
R_{eq}) = 0$] so that $\mq$ times the interior commutator gives
$[V_i^I(\tau) u_i(\tau),R_{eq} \hat{\rho}(\tau)]$. Similarly,
writing $\mq = 1-\mathcal{P}$ for the second from the interior
commutator, and simplifying, the bath trace operation gives us the
two time bath correlation functions $\la u_i (\tau) u_i(\tau_1)
\ra_{R}$. Repeating the same operations for the remaining
commutators, noting that the bath averaging for the odd bath
operators vanish and using the Wick theorem $\la u_i u_j u_k u_l
\ra_{R} = \la u_i u_j \ra_{R} \la u_k u_l \ra_{R} + \la u_i u_k
\ra_{R} \la u_j u_l \ra_{R} + \la u_i u_l \ra_{R} \la u_j u_k
\ra_{R}$, we obtain that the fourth order term includes the
product of two time bath correlation functions i.e., $\la u_i
(\tau) u_i(\tau_1) \ra_{R} \la u_j (\tau_1) u_j(\tau_2) \ra_{R}$.
On converting the equation into dimensionless time form as
described in Section IV, and using the high temperature
approximation for the correlation function, we conclude that the
order-of-magnitude of the fourth order term is;
\begin{equation}
\textrm{A}_4 \sim \frac{\lambda^2}{\gamma^4 \beta^2 \hbar^4}.
\end{equation}

The ratio $R_{42}$ of the fourth-order term to the second-order term is therefore
$R_{42} =\frac{\lambda}{\gamma^2 \beta \hbar^2}$, which is of the order of
$0.074$ for $\lambda = 1~\textrm{cm}^{-1}$, and $\sim 0.74 $ for
$\lambda = 10~\textrm{cm}^{-1}$. This suggests that the second
order approximation for the master equation is good for $\lambda
\sim 1$ for the standard set of parameters ($\gamma^{-1} = 100~\textrm{fs},~~J = 20 ~\textrm{cm}^{-1},~~
\Delta = 100~\textrm{cm}^{-1},~~T = 300~$K). However,
for large $\lambda \sim 10$ the fourth order term cannot be
neglected. Interestingly, the domain of applicability of the
second-order approximation in this $J \sim \Delta$ regime
has a dependence on the same collection of parameters,
$(\lambda/\gamma^2\beta\hbar^2$ small), as does
the Markov approximation in the $J \ll \Delta$.


\section{Initial state preparation by an ultra-short laser pulse}

\begin{figure}
\includegraphics[height = 6cm, width = 7cm]{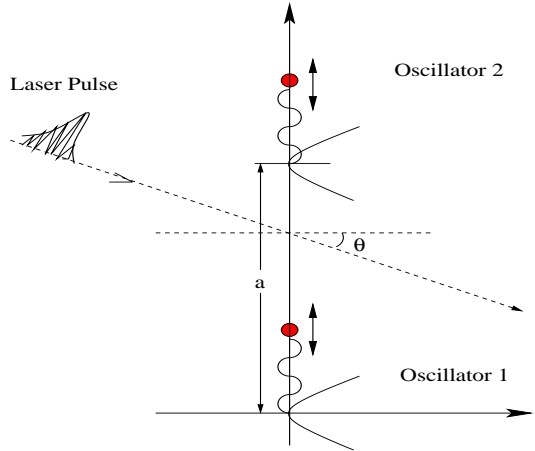}
\caption{Two harmonic oscillators excited by an ultra-short laser pulse.} \label{fig4}
\end{figure}

\subsection{Formulation}

To understand the effect of initial coherences on the subsequent
quantum dynamics, we consider a model of two 1-D harmonic oscillators
separated by a distance $a$ that are excited by an ultra short laser
pulse (Fig. \ref{fig4}). The results of this excitation are used below as
sample coherent initial conditions for dimer propagation. Note that we
restrict attention,  as do most treatments of excitation of light-harvesting
systems, to the ``one-exciton manifold", i.e. single excitations on each 
site. This model is useful to examine the dynamics, but should be augmented
by excitation of states with bi-excitons of examining issues like entanglement,
where the contributions from higher exciton states, no matter how small
 in magnitude, affect the entanglement measure.

In the system (Fig. \ref{fig4}) the wave vector of the laser pulse along the
propagation direction makes an angle $\theta$ with the line
perpendicular that joining the oscillators. The
laser field is treated semiclassically with the field
sufficiently weak to allow first order perturbation theory for the
light-oscillator interaction \cite{schiff}. The total
Hamiltonian in the coordinate representation is then
\begin{eqnarray}
 H_T& =& H_{sys} + H_{int}\nonumber\\
H_{sys}& =& -\frac{\hbar^2}{2 m}\frac{\pr^2}{\pr y_1^2}
 + \frac{1}{2} k_1 y_1^2 - \frac{\hbar^2}{2 m}\frac{\pr^2 }{\pr y_2^2} + \frac{1}{2} k_2 y_2^2\nonumber\\
H_{int}&= &\frac{i e \hbar}{2 m c} \bf{A}.\bf{\nabla}\nonumber\\
 {\bf{A}}. {\bf{\nabla}}& =& A_1(y_1,\theta,t)\hat{\ep}.
\hat{j}\frac{\pr}{\pr y_1} + A_2(y_2,\theta,a,t)\hat{\ep}.
\hat{j}\frac{\pr}{\pr y_2}~,
\end{eqnarray}
with the vector potential ${{\bf A}(y,\theta,t)} =
\sum_{{\bf{k}}} \hat{\ep}_{\bf{k}} (A_k e^{i (k y
\sin\theta-\omega t)} + A_k^{\ast} e^{-i (k y
\sin\theta-\omega t)} )$. For a coherent laser pulse,
$\hat{\ep}_{\bf k} = \hat{\ep}$.
If $A^\ast (-\frac{\omega}{c}) = A(\frac{\omega}{c})$, where $c$ is the speed
of light, 
then changing the summation to integration assuming continuous
distribution of modes, we have
\begin{equation}
{\bf A}(y,\theta,t) = \hat{\ep} \int_{-\infty}^{\infty} d\omega A(\omega/c) \exp[i \omega (y \sin(\theta)/c - t)] .
\end{equation}

For a Gaussian pulse $A(\omega/c) = A_0 \exp[-\eta (\omega -
\omega_0)^2]$, where $\omega_0$ is the central pulse frequency
and $\eta$ defines the pulse width. The vector potentials
take the form
\begin{eqnarray}
 A_1(y_1,\theta,t)= && A_0 \sqrt{\frac{\pi}{\eta}}~\exp\left[ i\omega_0 \left(\frac{y_1 \sin\theta}{c} - t \right)\right]
 \exp\left[-\frac{1}{4 \eta}\left(\frac{y_1 \sin\theta}{c} - t\right)^2\right]\nonumber\\
A_2(y_1,\theta,a,t)=&& A_0
\sqrt{\frac{\pi}{\eta}}\exp\left[i\omega_0 \left(\frac{y_2
\sin\theta}{c} - \frac{a}{c}\sin(\theta)- t
\right)\right]\nonumber \\ && \exp\left[-\frac{1}{4
\eta}\left(\frac{y_2 \sin\theta}{c}- \frac{a}{c}\sin(\theta) -
t\right)^2\right]
\end{eqnarray}
The laser frequency is assumed tuned so as to excite the first excited
state, with both oscillators initially (at $t = -\infty$)
in their ground states.

The eigensystems of oscillators 1 and 2 are
\begin{eqnarray}
&& E_n^{(1)} = \hbar \omega_{c1}(n + 1/2),~~ u_0^{(1)} = \frac{\sqrt{\alpha_1}}{\pi^{1/4}} \exp[-\frac{1}{2}\alpha_1^2 y_1^2],~~~ u_1^{(1)} = \frac{\sqrt{2}}{\pi^{1/4}} \alpha_1^{3/2} y_1 \exp[-\frac{1}{2}\alpha_1^2 x_1^2]\nonumber\\
&& E_n^{(2)} = \hbar \omega_{c2}(n + 1/2),~~ u_0^{(2)} =
\frac{\sqrt{\alpha_2}}{\pi^{1/4}} \exp[-\frac{1}{2}\alpha_2^2
y_2^2],~~~ u_1^{(2)} = \frac{\sqrt{2}}{\pi^{1/4}} \alpha_2^{3/2}
y_2 \exp[-\frac{1}{2}\alpha_2^2 y_2^2].
\end{eqnarray}
with $\omega_{ci} = \sqrt{k_i/m}$ and $\alpha_i^4 = m
k_i/\hbar^2,~~ i = 1,2$. The total wavefunction of the system is
\begin{equation}
\Psi(y_1,y_2,t)= \sum_{m,n} a_{mn}(t) u_n^{(1)}(y_1)u_m^{(2)}(y_2) \exp[-\frac{i}{\hbar}(E_n^{(1)} + E_m^{(2)})t],~~~~ a_{mn}(t = -\infty) = \delta_{n0}\delta_{m0}.
\end{equation}
Standard first-order perturbation theory gives the coefficients
as
\begin{eqnarray}
&& a_{nm}(t) = \frac{e \cos\theta}{2 m c}\int_{-\infty}^t dt'\int_{-\infty}^{+\infty}dx_1\int_{-\infty}^{+\infty}dx_2 u_n^{(1)}(x_1)u_m^{(2)}(x_2) \nonumber\\
&&\times\left( A_1(x_1,\theta,t') (\frac{\pr}{\pr x_1}) + A_2(x_2,\theta,a, t') (\frac{\pr}{\pr x_2}) \right)u_0^{(1)}(x_1)u_0^{(2)}(x_2) \exp[-i \omega_{nm}t']
\end{eqnarray}

Using the dipole approximation,
the spatial integrals for both $a_{01}$ and $a_{10}$ can be done
exactly. The remaining time integral is treated as follows: the laser
pulse is assumed to be ultrashort compared to the subsequent
quantum dynamics. For
times much greater than $t/\sqrt{\eta}$, the
exponential $e^{-(t^2/4 \eta)}$ in time integration will
be small, and the upper limit of the time
integration can be extended to $+\infty$. The integration can then
be performed exactly, giving
\begin{eqnarray}
&& a_{10} = -\zeta \sqrt{\mu} \cos\theta e^{- \eta \omega_0^2 (1+\nu_{10})^2}~~, \nonumber\\
&&a_{01} = -\zeta \frac{\cos\theta}{\sqrt{\mu}} e^{- \eta \omega_0^2 (1+\nu_{01})^2}.
\end{eqnarray}
Here $\mu = \alpha_1/\alpha_2 = (k_1/k_2)^{1/4},~~\zeta = \pi A_0
e \sqrt{\alpha_1 \alpha_2}/{\sqrt{2}} m c, ~~ \nu_{01} =
\omega_{01}/\omega_0,~~ \nu_{10} = \omega_{10}/\omega_0$, with
$\omega_{mn} = (E_0^{(1)} +E_0^{(2)} - E_m^{(1)} -
E_n^{(2)})/\hbar$.

The first excited states of the both oscillators
($E_1^{(1)},~~E_1^{(2)}$) constitute our relevant system (they are
separated by about 100 cm$^{-1}$ in typical systems like FMO), and
quantum dynamics takes place between them.  The
superposition of the excited states is written as
\begin{equation}
|\psi\ra = a_{10} |10\ra + a_{01} |01\ra,
\end{equation}
where $| 10\ra$ indicates that the first oscillator is excited and
the second is in the ground state. The density matrix at the
initial time is
\begin{equation}
\rho_0 = |\psi\ra\la \psi| = |a_{10}|^2 |10\ra\la10| + |a_{01}|^2 |01\ra\la01| + a_{10}a_{01}^\ast |10\ra\la01| + a_{01}a_{10}^\ast |01\ra\la10|.
\end{equation}

This initial density matrix corresponds to a particular orientation angle
$\theta$. For excitation of an ensemble we average over theta,
\begin{equation}
\la \rho_0\ra_\theta \equiv \frac{1}{2\pi}\int_0^{2\pi} \rho_0 d\theta,
\end{equation}
with the normalization
\begin{equation}
\rho_{11}+\rho_{22} \equiv \la |a_{10}|^2\ra_\theta + \la |a_{01}|^2\ra_\theta = 1.
\end{equation}
Thus,
\begin{eqnarray}
\rho_{11}& =& \frac{\la |a_{10}|^2\ra_\theta}{\la
|a_{10}|^2\ra_\theta + \la |a_{01}|^2\ra_\theta }\nonumber
\\ \rho_{22}& =& \frac{\la |a_{01}|^2\ra_\theta}{\la
|a_{10}|^2\ra_\theta + \la |a_{01}|^2\ra_\theta }\nonumber
\\ \rho_{12}&=& \rho_{21}^* = \frac{\la a_{10} a_{01}^\ast \ra_\theta}{\la
|a_{10}|^2\ra_\theta + \la |a_{01}|^2\ra_\theta }.
\label{eq39}
\end{eqnarray}
with
\begin{eqnarray}
\la |a_{10}|^2\ra_\theta &=& \frac{1}{2}\zeta^2 \mu e^{-2\eta \omega_0^2 (1+\nu_{10})^2},~~~\la |a_{01}|^2\ra_\theta= \frac{1}{2}\zeta^2 \frac{1}{\mu} e^{-2\eta \omega_0^2 (1+\nu_{01})^2}\nonumber\\
 \la a_{10} a_{01}^\ast \ra_\theta &=& \frac{c}{a
\omega_{01}}\zeta^2 J_1(\frac{a \omega_{01}}{c}) e^{-\eta
\omega_0^2 [(1+\nu_{10})^2 + (1+\nu_{01})^2]}.
\label{eq41}
\end{eqnarray}
Here $J_1(.)$ is the Bessel function of first kind and of order $1$. This constitutes
the initial density matrix for the relevant system.

\subsection{Numerical Results}

\begin{figure}
\centering
\begin{tabular}{cc}
\includegraphics[height = 4cm, width = 5cm]{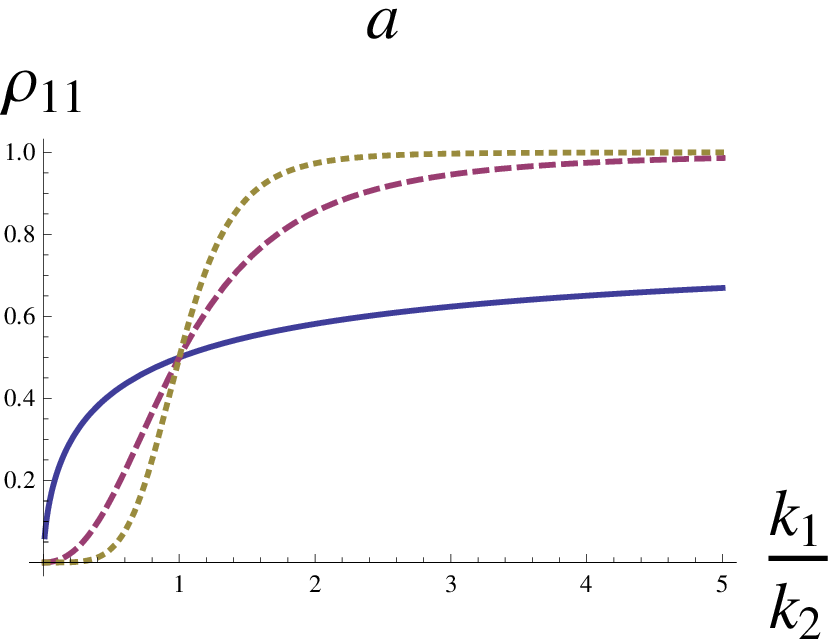}~~~~~~~&
\includegraphics[height = 4cm, width = 5cm]{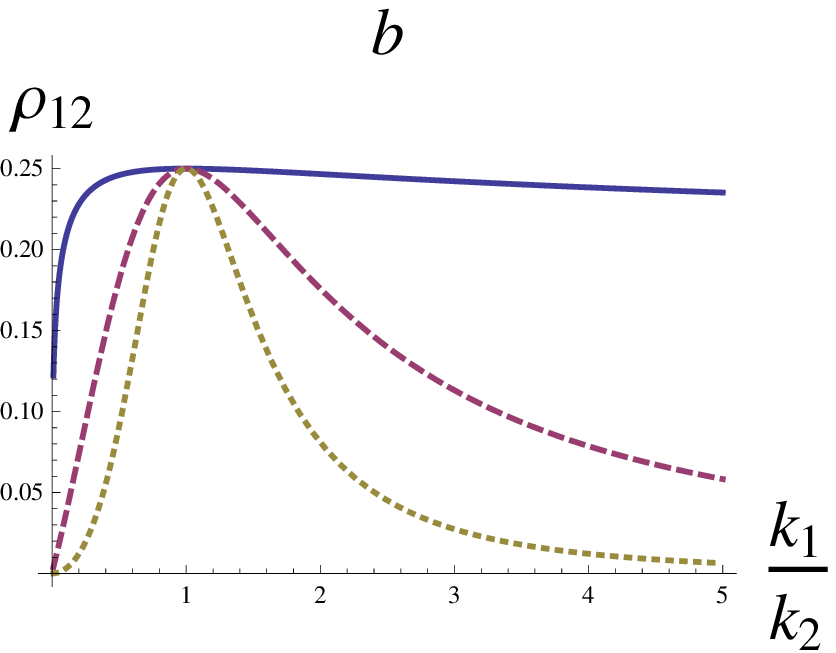}
\end{tabular}
\caption{(a) Population of angle averaged energy state $|10\ra\la 10|$, and (b) Coherence $\rho_{12}$ as a function of $\mu = \frac{k_1}{k_2}$. 
Solid line, laser frequency $\omega_0= 10^{14} $Hz, dashed line $5 \times 10^{14}$ Hz (Orange),
dotted line $10^{15}$ Hz. Note that Im$\rho_{12} =0$.} \label{fig5}
\end{figure}

Let $\eta = 10^{-4}$~ps$^2$, the separation between the two oscillators $a = 2$
~nm, ~$\omega_{01} = - \omega_{c2} =- 10^{14} $ Hz, and
$\omega_{10} = \sqrt{\mu}\omega_{01}$.
Figure \ref{fig5} shows the density matrix elements [i.e. the initial state given
by Eqs. (\ref{eq39}) and (\ref{eq41})] as a function of the
ratio of oscillator for constants
$k_1/k_2$ for various values of the excitation laser
frequency.

\begin{figure}
\centering
\begin{tabular}{ccc}
\includegraphics[height = 4cm, width = 5cm]{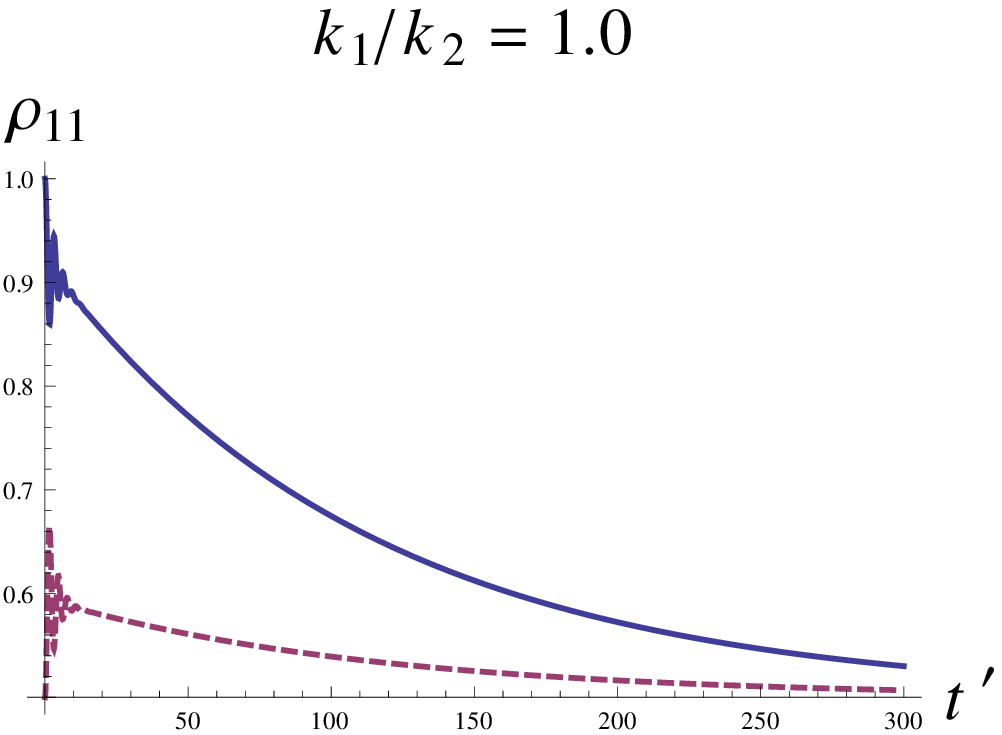}~~~~~~~&
\includegraphics[height = 4cm, width = 5cm]{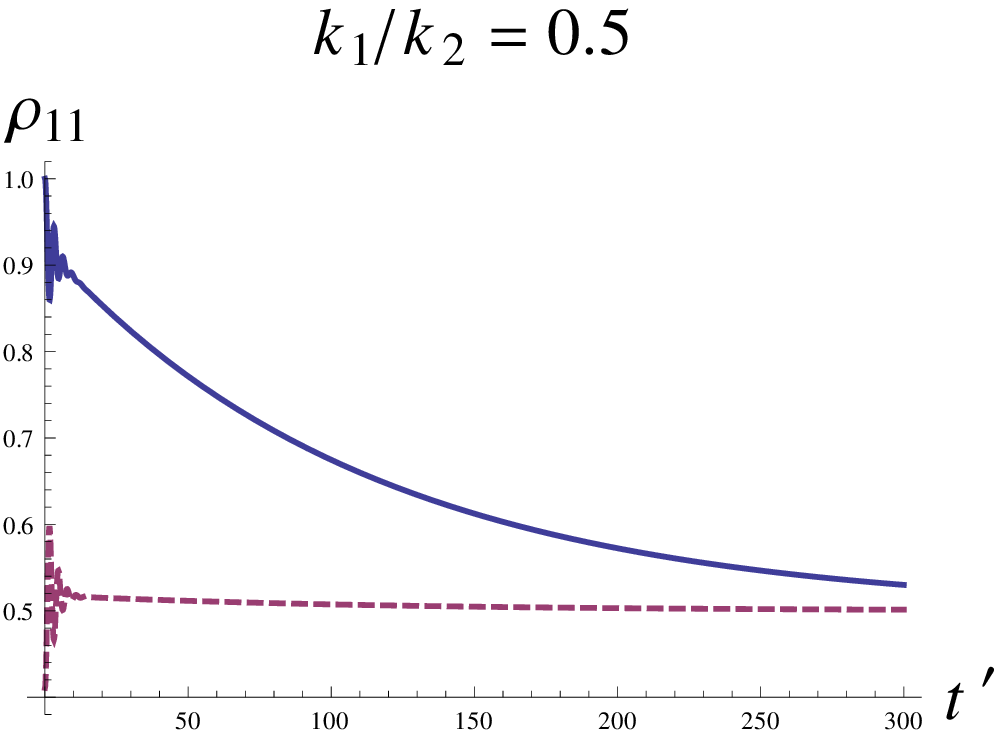}~~~~~~~&
\includegraphics[height = 4cm, width = 5cm]{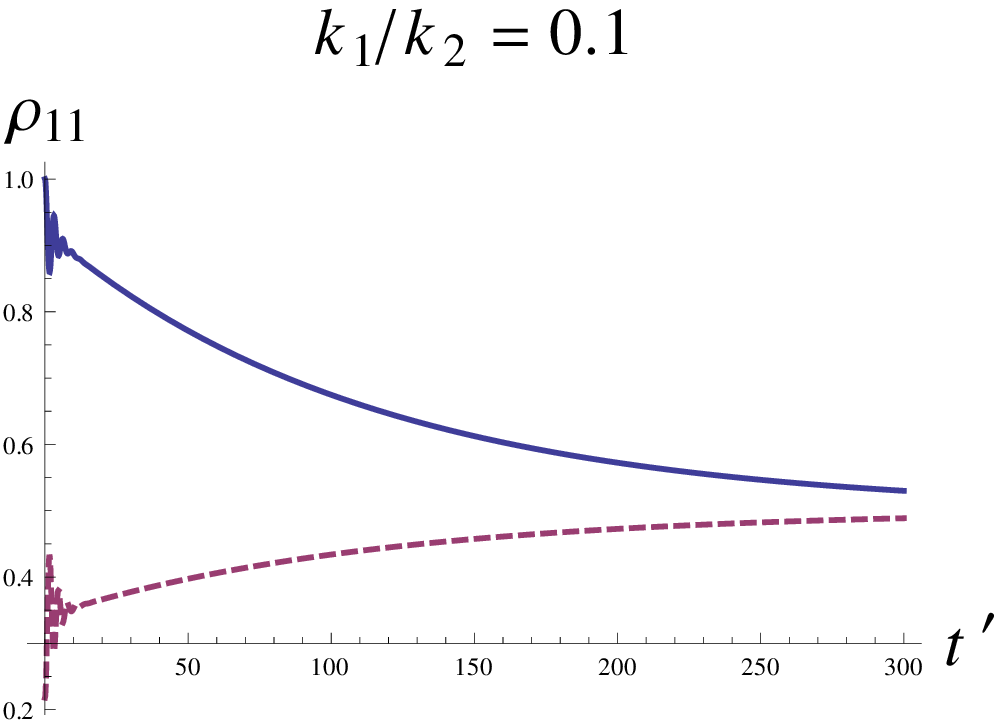}\\

\includegraphics[height = 4cm, width = 5cm]{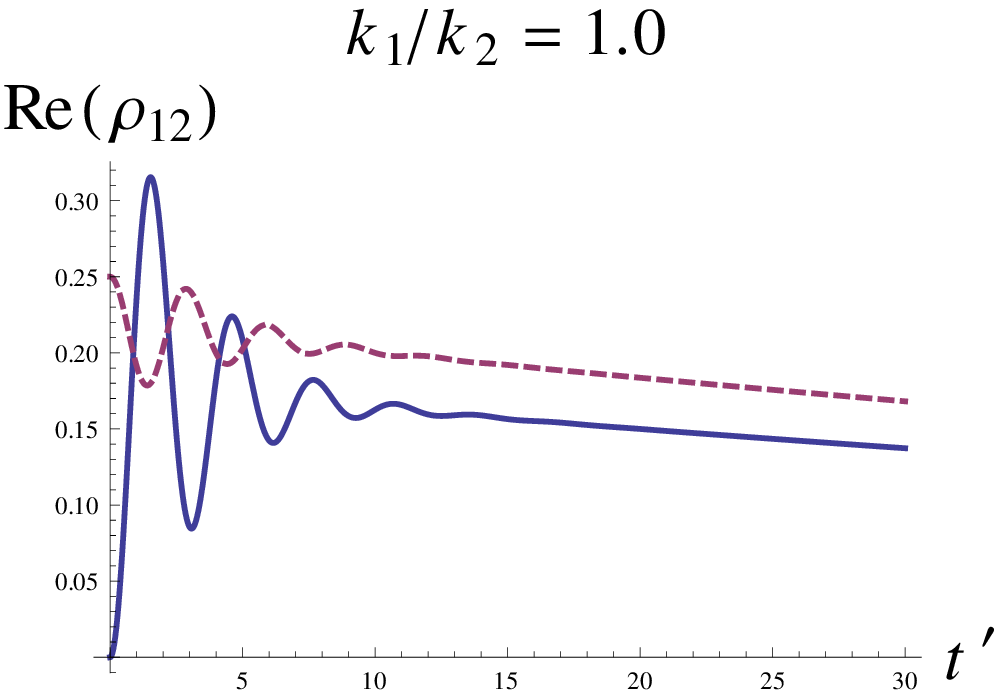}~~~~~~~&
\includegraphics[height = 4cm, width = 5cm]{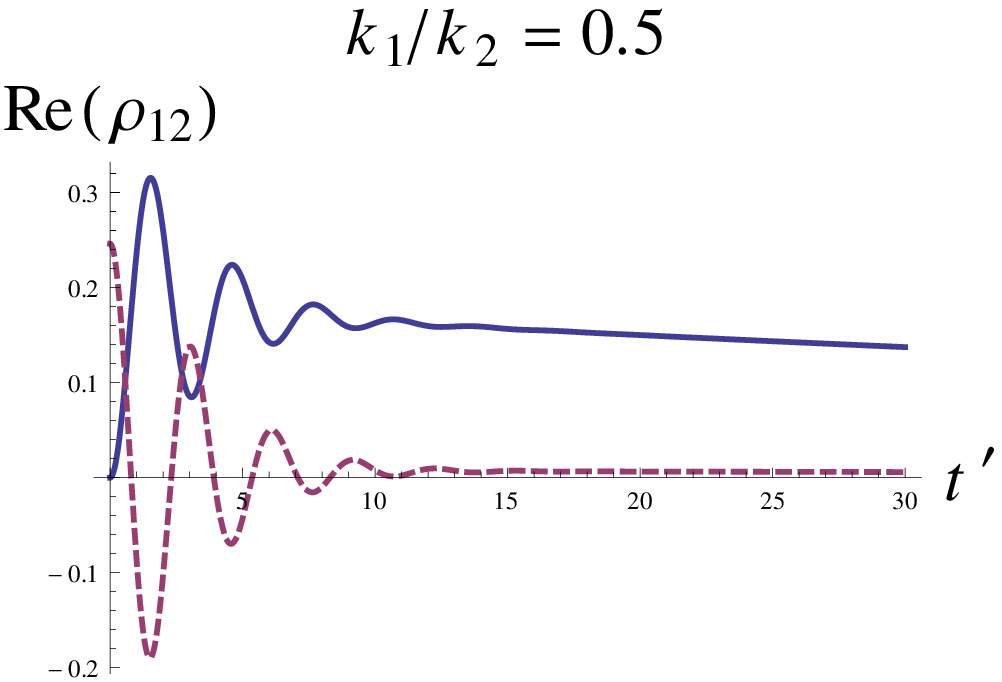}~~~~~~~&
\includegraphics[height = 4cm, width = 5cm]{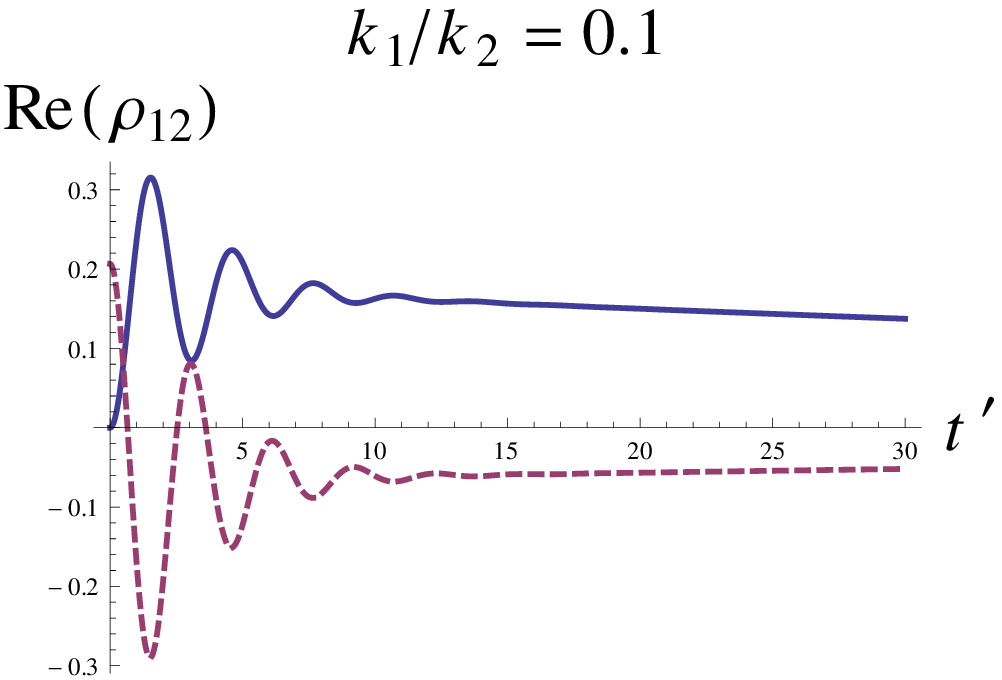}\\

\includegraphics[height = 4cm, width = 5cm]{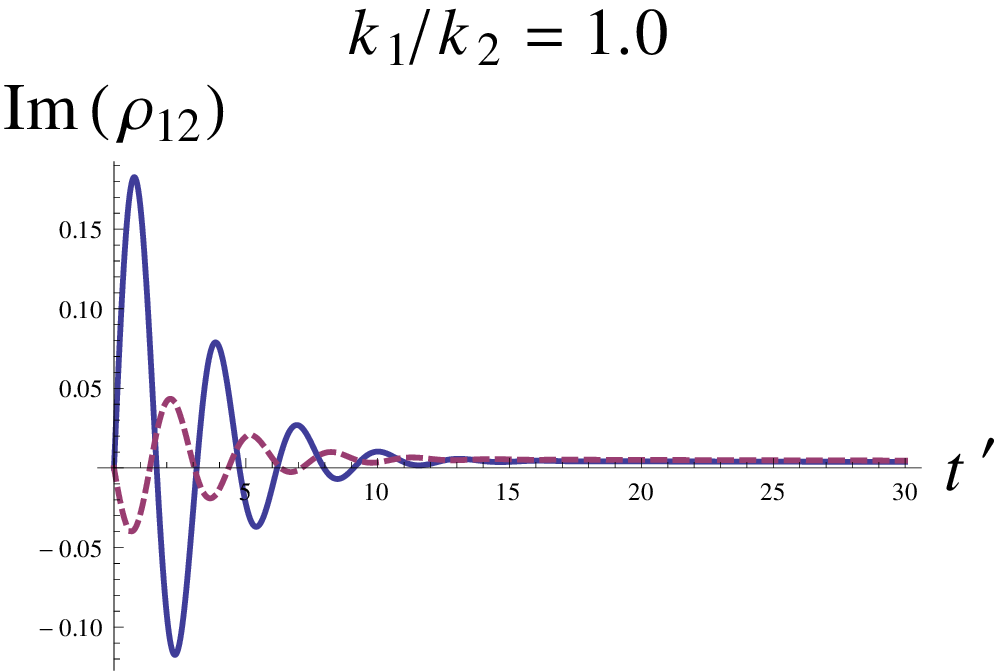}~~~~~~~&
\includegraphics[height = 4cm, width = 5cm]{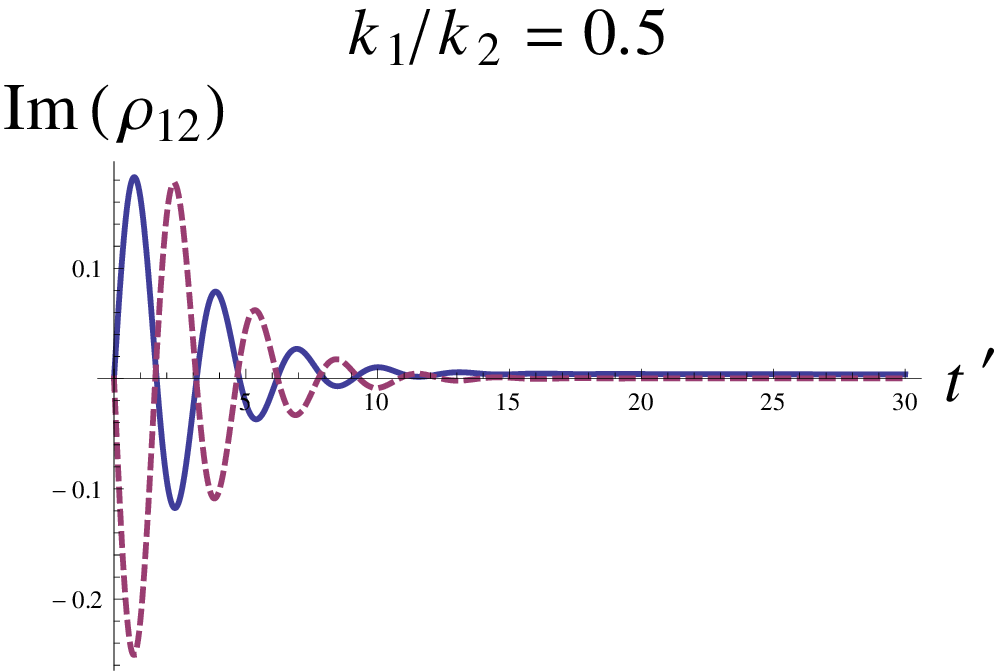}~~~~~~~&
\includegraphics[height = 4cm, width = 5cm]{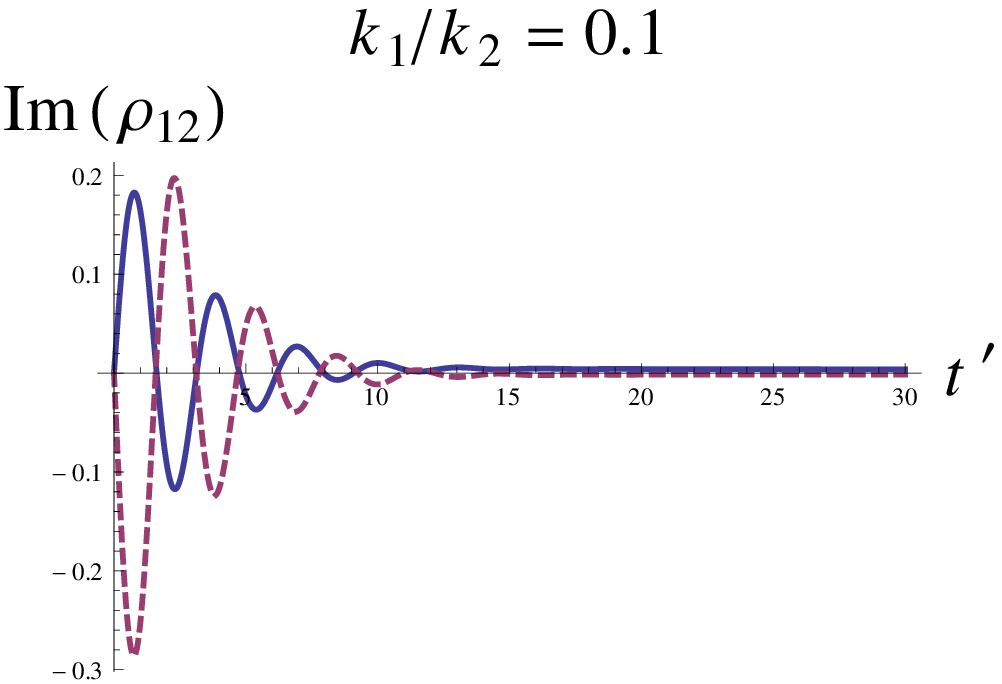}

\end{tabular}
\caption{ The effect of initial coherence on the
of relaxation dynamics. Long time behavior for various values of $k_1/k_2$. Here $\lambda = 1$ cm$^{-1}$ and
$\omega_0 = 10^{14}$. The blue (solid) curve is for
``no-coherence" initial condition , and red (dotted) curve is for
the ``initial coherence" condition, i.e., initial value of the
density matrix elements for the model photo-excitation studied
above. The second and third row shows the dynamics of the
off-diagonal elements.} \label{fig6}

\end{figure}

This initial state can then be used as a model for the initial conditions
for the numerical solution of the non-Markovian equations [Eq. (\ref{eq15})]
for the dimer. Figure
\ref{fig6} (in the site representation) compares the time evolution of
$\rho_{11}(t)$ for an initial ``no-coherence" state (only
populations) and the model generated state with ``initial coherence", for
various values of $k_1/k_2$.

One may identify two time scales associated with the electronic energy
transfer, the time scale over which the site occupation $\rho_{11}$ becomes
relatively constant, and the rate at which this occurs.
From the plots of $\rho_{11}$ in Fig. \ref{fig6}, it appears that
the presence of initial coherence ( at $t = 0$) effects both of these
time scales, but has little effect on the overall damping-out of the
coherence, i.e. the overall decay of oscillations in both the real and
imaginary parts of $\rho_{1,2}$. By contrast, the fall-off rate for the
decay of $\rho_{11}$ is far faster for the case with initially no-coherence
than it is for the case where there initially is coherence. In cases other
than $k_1/k_2=1.0$ the time at which the system reaches the equilibrium
value of 1/2 seems similar in both the cases where there is coherence
initially and where there is not.

\section{Conclusion}

A straightforward approach to 
solving the second order Born master equation, with and
without the Markov approximation, has been introduced. In addition to
obtaining numerical results showing the range of validity of these 
approximations, a number of 
analytical estimates, shown in Table I,  of parameter ranges over which these
approximations can be used has been obtained. For the case of the 
traditional dimer model for electronic energy transfer in photosynthesis,
surprisingly small reorganization energies (a few cm$^{-1}$) are required
for the validity of the Markovian approximation. In addition, we note that
for dimer coupling strengths on the order of the energy difference between 
site energies, higher order terms than second order in the system-bath
coupling are required if 
$4 \lambda/(\hbar^2 \beta \gamma^2) << 1$ is not satisfied,
where $\lambda$ is the reorganization
energy, and $\gamma$ defines the exponential 
falloff rate of the bath correlation function. Once again, the limitation to
small reorganization energies, not well appreciated in the past,
 is made explicit. 

We have also provided an example of the role of initial coherences in the
subsequent evolution of  the dimer dynamics for typical parameters 
associated with model photosynthetic light harvesting systems.

\section{Acknowledgment}
Financial support from the Natural Sciences and Engineering Research Council of Canada and from the U.S. Air Force Office of Scientific
Research under grant number FA9550-10-1-0260 is gratefully acknowledged.

\end{document}